\documentclass[11pt]{article}
\pdfoutput=1
\usepackage{jcapmod}
\usepackage{mathtools}
\usepackage{microtype}
\usepackage[utf8]{inputenc}
\usepackage[english]{babel}
\usepackage[dvipsnames]{xcolor}
\usepackage{amsmath,amssymb,amsbsy,amstext,amsthm}
\usepackage[colorlinks=true]{hyperref}
\usepackage{microtype}
\usepackage{graphicx}
\usepackage{amsfonts}
\usepackage{upgreek}
\usepackage{exscale,relsize}
\usepackage[makeroom]{cancel}
\usepackage{soul}
\usepackage{float}
\usepackage{empheq}
\usepackage[most]{tcolorbox}
\RequirePackage{color}
\usepackage{feynmp}
\usepackage{ifpdf}
\ifpdf
\fi

\usepackage{colortbl}
\definecolor{green2}{cmyk}{0, 1, 0.5, 0}
\definecolor{lightgreen}{cmyk}{0.2, 0, 0.2, 0.2}
\definecolor{dred}{rgb}{0.9,0.2,0.5}
\definecolor{dred2}{cmyk}{0.1,0.7,0.1,0.3}
\definecolor{lightgray2}{cmyk}{0.4,0.4,0,0.8}
\definecolor{black}{cmyk}{1.0,1.0,1.0,1.0}

\definecolor{verde}{rgb}{0,0.5,0}

\AtBeginDocument{
 \hypersetup{
 urlcolor=NavyBlue,
 citecolor=NavyBlue,
 linkcolor=BrickRed,
 }
}

\allowdisplaybreaks[1]

\usepackage{colortbl}

\makeatletter
\newlength{\apb@width}
\newcommand{\autoparbox}[2][c]{\settowidth{\apb@width}{#2}\parbox[#1]{\apb@width}{#2}}

\makeatother

\numberwithin{equation}{section}

\def\beq{\begin{equation}}
\def\eeq{\end{equation}}

\def\bea{\begin{eqnarray}}
\def\eea{\end{eqnarray}}
\def\eg{{\it e.g.~}}

\def\ie{{\it i.e.~}}
\def\d{{\rm d}}

\def\d{{\rm d}}

\def\nn{\nonumber}

\def\fr{\frac}

\def\0{{\boldsymbol 0}}

\def\fr{\frac}

\newtcbox{\mymath}[1][]{%
    nobeforeafter, math upper, tcbox raise base,
    enhanced, colframe=gray!30!gray,
    colback=gray!10, boxrule=0.5pt,
    #1}

\usepackage{setspace} 

\begin{document}

\begin{titlepage}

\setcounter{page}{1} \baselineskip=15.5pt \thispagestyle{empty}

\bigskip\

\vspace{1cm}
\begin{center}

{\fontsize{19}{28}\selectfont  
{\bf A universal constraint on axion non-Abelian dynamics\vspace{0.35cm}\\ during inflation
}}

\end{center}

\vspace{0.2cm}
\begin{center}
{\fontsize{13}{30}\selectfont Ema Dimastrogiovanni$^{\dagger}$,
}
{\fontsize{13}{30}\selectfont Matteo Fasiello$^{\ddagger}$,
}
{\fontsize{13}{30}\selectfont Martino Michelotti$^{\dagger}$
}and
{\fontsize{13}{30}\selectfont Ogan \"Ozsoy$^{\ddagger}$
}
\end{center}
\begin{center}
\textsl{$^\dagger$ Van Swinderen Institute for Particle Physics and Gravity, University of Groningen, Nijenborgh
4, 9747 AG Groningen, The Netherlands.}
\vskip 2pt
\textsl{$^\ddagger$ Instituto de Física Téorica UAM-CSIC, c/ Nicolás Cabrera 13-15,
28049, Madrid, Spain.}
\vskip 8pt
\end{center}

\vspace{1.2cm}

\noindent
\begin{abstract}

Inflationary models equipped with Chern-Simons coupling between their axion and gauge sectors exhibit an array of interesting signals including a testable chiral gravitational wave spectrum. The energy injection in the gauge sector triggered by the rolling axion leads to a well-studied enhancement of gauge field fluctuations. These may in turn affect observables such as the scalar and tensor spectra and also account for non-linear corrections to field propagators. In this work, we focus on non-Abelian gauge sectors. We show that gauge field self-interactions and axion-gauge field non-linear couplings significantly renormalize the gauge field mode function. Operating within the regime of validity of the perturbative treatment places strong constraints on the accessible parameter space of this class of models. We calculate corrections to the gauge field propagator that are universally present in these scenarios. Enforcing perturbativity on such propagators leads to bounds that are competitive with those stemming from analytical estimates on the onset of the strong backreaction regime. 
\end{abstract}
\vspace{0.6cm}
 \end{titlepage}

\tableofcontents

\section{Introduction}

 Perhaps the simplest explanation of cosmological puzzles \cite{Weinberg:2008zzc} such as the horizon problem is to postulate a phase of accelerated expansion in the very early universe, cosmic inflation. The minimal inflationary scenario consists of a single scalar field slowly rolling (SFSR) down its potential. Several single-field models such as Higgs \cite{Bezrukov:2007ep} and Starobinsky \cite{Starobinsky:1980te} inflation, fiber inflation \cite{Cicoli:2008gp}, as well $\alpha$-attractors \cite{Kallosh:2013hoa}\footnote{Starobinsky inflation is a special case of the larger $\alpha$-attractor class.} remain in excellent agreement with CMB observations \cite{Planck:2018jri,BICEP:2021xfz}. One would like to ask (at least) one more property of the inflationary Lagrangian, namely that the inflaton mass be naturally light. This requirement is clearly not satisfied by generic inflaton potentials\footnote{This is known as the inflationary $\eta$ problem.} but rather is typically associated with the presence of a symmetry protecting the inflaton mass from large quantum corrections.
 
Axion inflation scenarios \cite{Pajer:2013fsa} enjoy an approximate shift symmetry and are therefore the poster child for natural models supporting early-time acceleration. The natural inflation model of \cite{Freese:1990rb} is the progenitor for the entire axion inflation class and, albeit ruled out, still serves as an important proxy to model the behaviour of inflation via axion-like-particles (ALPs). The standard cosine potential of the axion-inflaton is too steep \cite{Freese:2014nla} to grant agreement with experimental data. As a result, a variety of mechanisms have been explored to flatten the potential and/or slow down the inflaton rolling \cite{Kim:2004rp,Anber:2009ua,Barnaby:2011vw}, ideally without advocating a trans-Planckian axion decay constant. One such possibility relies on the friction obtained by coupling the axion to a gauge sector via Chern-Simons coupling \cite{Anber:2009ua}. \\
  \indent This last class of models has been the subject of an intense research activity \cite{Anber:2009ua,Barnaby:2010vf,Cook:2011hg,Maleknejad:2011sq,Barnaby:2011vw,Adshead:2012kp,Dimastrogiovanni:2012st,Mukohyama:2014gba, Ozsoy:2014sba,Namba:2015gja,Peloso:2016gqs,Obata:2016tmo,Dimastrogiovanni:2016fuu,Adshead:2016omu,Ozsoy:2017blg,Agrawal:2017awz,Caldwell:2017chz,Thorne:2017jft,Lozanov:2018kpk,Watanabe:2020ctz,Holland:2020jdh,Domcke:2020zez, Ozsoy:2021onx, Ishiwata:2021yne, Iarygina:2023mtj,Dimastrogiovanni:2023oid,Durrer:2024ibi,Putti:2024uyr,Dimastrogiovanni:2024xvc,Alam:2024krt,Lorenzoni:2024krn,Ozsoy:2024apn}. 
  Given the possibilities afforded by Abelian and non-Abelian scenarios, the chance to enter the strong backreaction\footnote{Backreaction in this context refers to the non-trivial effect of gauge field perturbations on the background equations of motion.} regime, and the number of degrees of freedom in play, it is no surprise that these scenarios exhibit an intriguing phenomenology spanning from a very distinctive gravitational wave (GW) spectrum (see e.g. \cite{Barnaby:2012xt,Mukohyama:2014gba,Thorne:2017jft,Agrawal:2018mrg,Fujita:2018vmv,Dimastrogiovanni:2018xnn,Domcke:2020zez,Ozsoy:2020ccy, Garcia-Bellido:2023ser}) to primordial black hole production \cite{Linde:2012bt,Lin:2012gs, Bugaev:2013fya, Erfani:2015rqv, Garcia-Bellido:2016dkw, Cheng:2016qzb, Cheng:2018yyr, Almeida:2020kaq, Ozsoy:2020kat, Ozsoy:2023ryl, Dimastrogiovanni:2024xvc}. The most striking features of this set of models can be traced back to the Chern-Simons (CS) coupling. This parity breaking term is behind the possibility of a chiral GW spectrum. Crucially, chirality is eminently testable in this context. At large scales one employs $\langle BT \rangle$ and $\langle EB \rangle$ correlations of CMB modes \cite{Lue:1998mq,Saito:2007kt,Gluscevic:2010vv, Gerbino:2016mqb}. At smaller scales both (a combination of) terrestrial \cite{Seto:2007tn,Smith:2016jqs} and space-based \cite{Domcke:2019zls} interferometers are sensitive to circularly polarized gravitational waves. Depending on the inflaton/spectator role played by ALPs, the dissipation of kinetic energy into the gauge sector can result into blue or bump-like scalar and tensor spectra. A large number of (necessarily spectator) axions may give rise to an entire GW forest \cite{Kitajima:2018zco,Dimastrogiovanni:2023juq}, a fraction of which could well be within reach for pulsar timing arrays as well as laser interferometers.
  
One should stress two aspects of the perturbative treatment of model that exhibit axion - gauge field dynamics. First, gauge field perturbation may backreact on the background equations of motion and significantly change the evolution. This was first and most prominently seen in the Abelian case \cite{Anber:2009ua} where the friction due to the CS coupling results in the production of gauge quanta. It is the non-trivial backreaction contribution that serves as friction. In the non-Abelian case there is room for significantly slowing down the inflaton/axion in the small backreaction regime, yet one would certainly want to access also the strong backreaction domain. Analytical approximations to tackle backreaction have been considered in \cite{Ishiwata:2021yne}, and an increasing number of lattice studies are becoming available \cite{Caravano:2021bfn,Figueroa:2023oxc}. The vast majority of these studies are on Abelian configurations, whilst the very recent interesting work in \cite{Iarygina:2023mtj} takes on the non-Abelian case \cite{Dimastrogiovanni:2016fuu}.\\
\indent Another key aspect lies in the  perturbativity limits on cosmological correlation functions. Standard observables are derived under the assumptions that loop effects will provide small controllable corrections. Explicitly imposing such consistency conditions translates into powerful bounds on the accessible parameter space of entire classes of inflationary theories. We shall focus here on ALPs coupled to a non-Abelian gauge sector and, much in the spirit of \cite{Ferreira:2015omg}, derive explicit bounds on the parameter space stemming from perturbativity limits.
In particular, we identify a correction to the propagator of the tensor modes $T$ in the gauge sector, arising from the gauge field self-interaction, that is always present in non-Abelian theories. This is a \emph{universal} correction in the sense that it is largely insensitive to the inflaton/spectator nature of the ALPs, as well as its coupling (minimal or otherwise) to gravity. In addition to this, we also derive the one-loop correction to the  $T$-propagator due to a scalar-tensor-tensor interaction  to impose additional bounds, but stress here that these are model-dependent constraints.

This paper is organized as follows. In \emph{Section} \ref{sec2}, we review the background and tensor perturbation dynamics of models featuring an axion (inflaton or spectator) coupled to non-Abelian gauge fields via a Chern-Simons interaction. In {\it Section} \ref{sec3}, we present one-loop correction for the two-point correlator of the gauge tensor perturbations;  highlighting how our findings provide novel perturbativity constraints for this class of models. We draw our conclusions in {\it Section} \ref{sec4}. {\it Appendices} \ref{AppA} and \ref{AppB} contain additional details on, respectively, the calculation of the perturbative correction and the dynamics of scalar perturbations in the specific setup we consider.

\section{Axion and non-Abelian gauge field dynamics during inflation}\label{sec2}

We will consider inflationary models with a non-Abelian ${\rm SU}(2)$ gauge field sector coupled via a Chern-Simons interaction with an axion-like field $\chi$. A useful starting point to discuss the typical dynamics of such a system is given by the following Lagrangian
\beq\label{LAGF}
\frac{\mathcal{L}_{\rm AGF}}{\sqrt{-g}}= -\frac{1}{4} F^{a \mu \nu} F_{\mu \nu}^a - \frac{\lambda\, \chi}{8 f \sqrt{-g}} \,\epsilon^{\mu \nu \rho \sigma} F_{\mu \nu}^a F_{\rho \sigma}^a \; .
\eeq
In the above expression, $(-g)$ is the determinant of the metric, not to be confused with the SU(2) gauge coupling $g$ appearing in the rest of the paper. Consistently with the (approximate) shift symmetry, the axion has a symmetry-preserving dimension-five interaction with the ${\rm SU}(2)$ gauge field $A^a_\mu$, whose strength is characterized by the dimensionless parameter $\lambda$ and the axion decay constant $f$: 
\beq\label{Lint}
\mathcal{L}_{\rm int} = - \frac{\lambda\, \chi}{8 f} \,\epsilon^{\mu \nu \rho \sigma} F_{\mu \nu}^a F_{\rho \sigma}^a\,.
\eeq
Here $\epsilon^{\mu\nu\rho\sigma}$ is the totally anti-symmetric tensor normalized as $\epsilon^{0123} = 1$. The field strength is defined by
\beq \label{gaugedef}
F_{\mu \nu}^a=\partial_\mu A_\nu^a-\partial_\nu A_\mu^a+g\, \epsilon^{a b c} A_\mu^b A_\nu^c\,,
\eeq
with $g$ denoting the gauge coupling and $\epsilon^{abc}$ the structure constants of the ${\rm SU}(2)$ algebra. \\

In order to emphasize the broad implications of the results obtained in this work, we stress that they are largely independent from the details of the axion Lagrangian, including any non-minimal coupling $\chi$ may have with gravity. Similarly, our arguments below do not depend on whether the axion-like field belongs to a hidden sector or takes an active role for the accelerated expansion. In particular, as we will show, as long as the axion-gauge field system is on a slow-roll attractor, an important relation that ties the background dynamics of the gauge and axion fields  can be derived. This relation in turn allows one to capture the underlying particle production in the  gauge sector in terms of a single parameter (commonly referred as $m_Q$ in the literature), which will be central for the universal constraints we shall derive. 
\medskip

\noindent{\bf Background evolution of the gauge field vacuum expectation value.} For this purpose, we consider the standard isotropic gauge field configuration at the background level:
\beq\label{gfb}
\langle A^{a}_0 \rangle (t) = 0, \quad\quad \langle A^{a}_i \rangle (t) = \delta_{i}^{a}\, a(t) Q(t),
\eeq
where $ a = 1,2,3$ is the gauge group index and the indices $0$ and $i = 1,2,3$ indicate time and space respectively. The Lagrangian \eqref{LAGF} in combination with the ansatz \eqref{gfb} give the well-known equation of motion for the gauge field vacuum expectation value (vev) $Q$ \cite{Adshead:2012kp}:
\begin{align}\label{Qeq}
    \ddot{Q}+3 H \dot{Q}+2 H^2 Q + 2 g^2 Q^3=\frac{\lambda g}{f} \dot{\chi} Q^2\,.
\end{align}
Assuming a slow-roll background trajectory $\ddot{Q} \ll 3 H \dot{Q}$, it is possible to find a non-trivial stationary solution for $Q$, identifying an effective potential in \eqref{Qeq} as
\beq\label{Qpot}
U_{\mathrm{eff}}(Q)\equiv H^2Q^{2}+\frac{1}{2}g^{2}Q^{4}-\frac{g\lambda}{3f}\dot{\chi}Q^{3}\,.
\eeq
Without loss of generality, we take $\dot{\chi} > 0$. As a result, in the slow-roll limit of \eqref{Qeq} the potential \eqref{Qpot} admits a non-trivial (positive) minimum for $Q$ for a non-vanishing axion velocity:
\beq\label{Qmin}
Q_{\rm min}=\frac{\lambda\dot{\chi}}{4gf}\left(1+\sqrt{1-\frac{16 f^{2}H^{2}}{\lambda^{2}\dot{\chi}^2}}\right)\ ,
\eeq
where for the reality of $Q_{\rm min} > 0$ we require $\lambda \dot{\chi} > 4fH$.
As long as the gauge field vev tracks this instantaneous minimum during inflation, its companion axion vev locks into the following slow-roll trajectory 
\beq\label{theR}
\dot{\chi}\simeq \frac{2fH}{\lambda}\left({\frac{gQ}{H}}+\frac{H}{gQ}\right)\ ,
\eeq
where we solved the stationarity condition $U_{\rm eff}'(Q) = 0$ for $\dot{\chi}$. 
On the slow-roll trajectory of the axion-gauge field system, Eq.~\eqref{theR} gives rise to a special relation that is often encountered in the literature on axion non-Abelian gauge field dynamics 
\beq\label{theRf}
\xi \simeq m_Q + m_Q^{-1}, 
\eeq
where we defined $m_Q \equiv {gQ}/{H}$ and $\xi \equiv \lambda\dot{\chi}/(2Hf)$. 

We would like to emphasize that in the steps  leading to Eq.~\eqref{theRf} one does not need to specify the form of the Lagrangian for the axion; the only assumption in place is for the axion-gauge field system to be in the slow-roll attractor solution described by Eqs.~\eqref{Qmin} and \eqref{theR}. This goes to show that Eq.~\eqref{theRf} is valid for a broad class of models characterised by a slow-roll dynamics of the axion-gauge field system during inflation. Such models include for example the spectator chromo-natural inflation model \cite{Dimastrogiovanni:2016fuu}, kinetically-driven inflation \cite{watanabe2020gravitational}, the recently proposed non-minimally coupled chromo-natural inflation \cite{Dimastrogiovanni:2023oid} and chromo-natural-like models with multiple inflation stages as in \cite{Fujita:2022jkc}.
\medskip

\noindent{\bf Gauge field production.} For the class of models we just listed, the interactions in the gauge sector, and therefore the dynamics of the ${\rm SU(2)}$ field, are  identical to those of chromo-natural inflation \cite{Dimastrogiovanni:2012ew,Adshead:2013nka}, as long as the background settles into the slow-roll attractor solution. The tensorial component of the non-Abelian gauge field, $T_{ai}\subset \delta A^a_i$, is known to exhibit a transient instability due to its coupling with the axion vev in Eq.~\eqref{Lint}. At the linearized level, the two polarization modes of the canonically-normalized, transverse and traceless ${\rm SU(2)}$ gauge tensor obey the following evolution equation \cite{Adshead:2013qp},
\beq\label{Teq}
\fr{\d}{\d x^2} T_{\pm}  + \left[1 + \frac{2 m_Q \xi}{x^2}\mp \frac{2(\xi+m_Q)}{x} \right] T_{\pm}  = \mathcal{O}(h_{\pm}), \quad\quad x \equiv  -k\tau = \frac{k}{aH},
\eeq
where T is expanded in Fourier space as
\beq\label{tft}
T_{ab}(\tau,\vec{x}) = \int \fr{\d^3 q}{(2\pi)^{3/2}}\, e^{i\vec{q}.\vec{x}} \sum_{\lambda = \pm} \Pi^{*}_{ab,\lambda}(\vec{q})\, T_{\lambda} (\tau,\vec{q}),
\eeq
with
\beq\label{Piab}
\Pi^*_{ab,\lambda}(\vec{k})=\epsilon_{a,\lambda}(\vec{k})\,\epsilon_{b,\lambda}(\vec{k}),
\eeq
and the polarization vectors satisfy the following relations 
\begin{align}\label{pve}
\nn & \vec{k} \cdot \vec{\epsilon}_{(\pm)}(\vec{k}) =0, \quad  \vec{k} \times  \,\vec{\epsilon}_{(\pm)}(\vec{k})=\mp i k  \, \vec{\epsilon}_{(\pm)}(\vec{k}), \\ &\vec{\epsilon}_{(\lambda)}(\vec{k})\cdot\vec{\epsilon}_{(\lambda^{\prime})}(\vec{k})^* =\delta_{\lambda \lambda^{\prime}},\quad \vec{\epsilon}_{(\pm)}(\vec{k})^*=\vec{\epsilon}_{(\pm)}(-\vec{k})=\vec{\epsilon}_{(\mp)}(\vec{k}).
\end{align}
On the right-hand-side of Eq.~\eqref{Teq}, we denoted with $\mathcal{O}(h_{\pm})$ the terms that arise from the mixing with the metric fluctuations $h_{\pm}$. Note that the mixing between the metric and $T_{\pm}$ is also responsible for generating  metric tensor perturbations, a contribution that is sensitive to the transient tachyonic effective mass exhibited by the $T_{+}$ modes around horizon crossing \cite{Adshead:2013qp}. The resulting amplification in the metric perturbation $h_+$ is not as strong as the corresponding enhancement in the $T_+$ modes. Moreover, the influence of $h_+$ on the evolution of the gauge tensor modes becomes relevant only sufficiently outside the horizon, $x = k/(aH) \ll 1$ \cite{Adshead:2013nka,Maleknejad:2016qjz,Dimastrogiovanni:2016fuu}. All in all, one can expect the mixing to amount to a negligible correction to the $T_+$ mode function \cite{Dimastrogiovanni:2016fuu}. In the interest of performing semi-analytical calculations, we focus on the contributions to non-linearities from around horizon crossing, where the gauge field is maximally enhanced.

As we emphasized earlier, an important property for all class of models we consider here is the fact that, along the inflationary trajectory, the relation \eqref{theRf} holds at leading order in slow-roll expansion (in the weak backreaction regime). 
This allows for the study of the dynamics of the dominant $T_{+}$ mode in terms of a single parameter, $m_Q$:
\beq\label{teq}
\fr{\d}{\d x^2} T_{\pm}  + \left[1 + \frac{2(1 + m_Q^2)}{x^2}\mp \frac{2(2m_Q+{m_Q}^{-1})}{x} \right] T_{\pm}  = 0.
\eeq
From \eqref{teq} one can easily verify that $T_+$ acquires a tachyonic mass for $x^{-}_{\rm th} < x < x^{+}_{\rm th}$, where we defined  
\beq\label{xinsta}
 x^{\pm}_{\rm th} \equiv 2 m_Q+{m_Q}^{-1}\pm\sqrt{2 m_Q^2+2+m_Q^{-2}},
\eeq
\begin{figure}[t!]
\begin{center}
\includegraphics[scale=0.475]{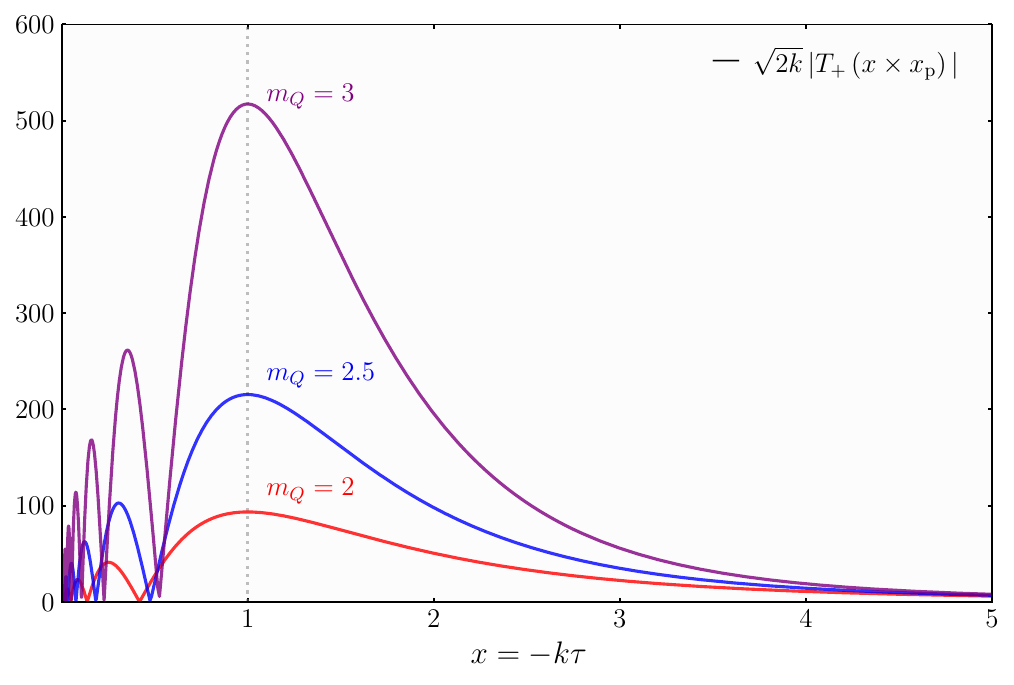}\includegraphics[scale=0.475]{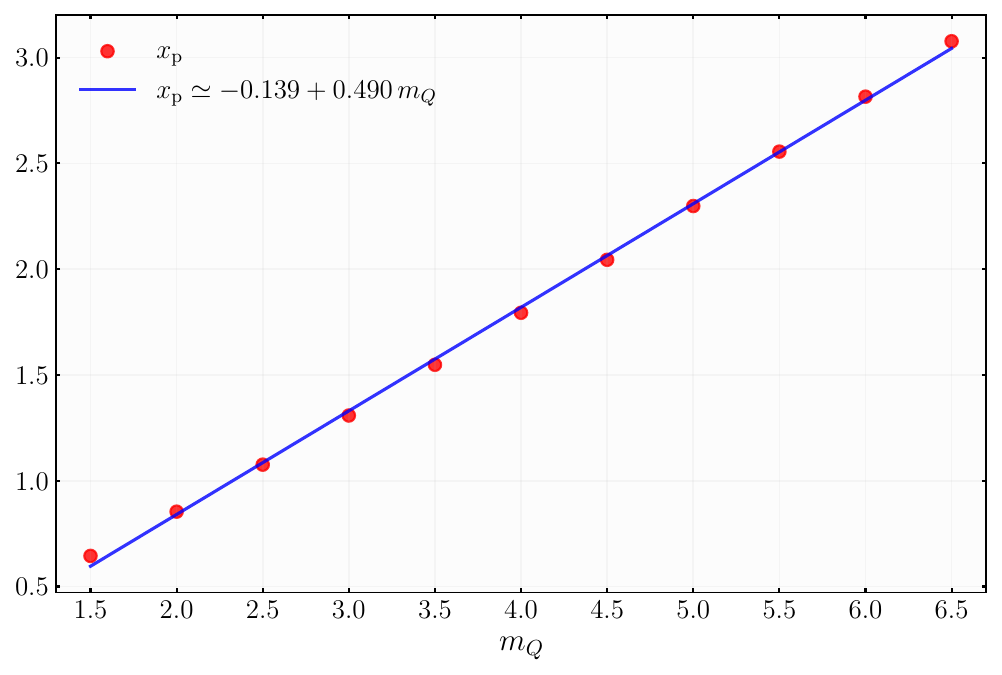}
\end{center}
\vspace*{-5mm}\caption{The amplification of the $T_+$ modes (left plot) for different $m_Q$ values as they evolve from inside $x > 1$ to outside $x < 1$ the Hubble horizon. The argument of the modes shown are scaled with $x_{\rm p}$, denoting the peak location of $|T_+|$ (for a given $m_Q$) that is obtained numerically. The accuracy of the analytic fitting form for the peak location is provided in the right panel.\label{fig:Tp}}
\end{figure}
with $(+)$ and $(-)$ denoting, respectively, the boundaries of the time frame during which the instability unfolds. Treating $m_Q$ as adiabatically constant in the slow-roll regime\footnote{In order to study the spectral dependence of quantities that depend on this parameter, one can evaluate $m_Q$ at the scale a given mode leaves the horizon as it is customarily done in models of slow-roll inflation.}, \textit{i.e.} $\dot{m}_Q/(Hm_Q)\ll1$, an explicit analytic solution for Eq.~\eqref{teq} can be derived to describe the amplification in the $T_+$ modes \cite{Adshead:2013qp,Obata:2016tmo,Dimastrogiovanni:2016fuu,Agrawal:2017awz},
\beq\label{tpk}
{T}_{+}(\tau, k) = \fr{1}{\sqrt{2k}}\,{\rm e}^{{\pi}(2m_Q + m_Q^{-1})/2}\,W_{\alpha, \beta} (2 i k \tau), \quad \alpha \equiv -i(2m_Q + m_Q^{-1}),\quad \beta \equiv -i \sqrt{2m_Q^2 + 7/4}\,.
\eeq
Here $W_{\alpha,\beta}$ is the Whittaker function and the exponential pre-factor ${\rm e}^{\pi(2m_Q + m_Q^{-1})/2}$ characterizes the enhancement induced by the instability in the typical regime of interest, where $m_Q \simeq \mathcal{O}(1)$. In Figure \ref{fig:Tp}, we present the amplification of the gauge tensor modes \eqref{tpk} (for different $m_Q$ values) as they evolve from sub-horizon to super-horizon scales. For illustrative purposes, we re-scaled the argument of the mode functions by the ($m_Q$ dependent) peak location $x_{\rm p}$, obtained numerically such that peaks of $T_+$ for different $m_Q$ values occur at the same location in the plot. The right panel in Figure \ref{fig:Tp} shows the accuracy of the fitting function (for $1.5 < m_Q < 6.5$) that locates the peak for a given choice of $m_Q$: 
\beq\label{xp}
x_{\rm p} \simeq -0.171 + 0.496\, m_Q.
\eeq
This expression will be instrumental for the study of non-linearities as we expect their influence to be significant at the peak of the particle production process. 

The enhancement experienced by the $T_+$ mode \eqref{tpk} is the starting point of many phenomenological applications found in the literature on axion - gauge field dynamics. These include sourced gravitational waves at CMB \cite{Adshead:2013qp,Dimastrogiovanni:2016fuu} and interferometer scales \cite{Thorne:2017jft,Campeti:2020xwn}, generation of sizeable scalar and tensor non-Gaussianities \cite{Agrawal:2017awz, Agrawal:2018mrg, Fujita:2018vmv, Dimastrogiovanni:2018xnn}, and of primordial black holes \cite{Dimastrogiovanni:2024xvc}. Central to the discussion for any inflationary model that exhibits non-Abelian gauge field production by the slowly-rolling ALPs, these approaches typically assume that the tensor fluctuations in the gauge sector remain in the perturbative regime. 
In this work, our aim is to assess the validity of this assumption by studying the non-linearities involved in the dynamics of $T_+$ modes. In the following, we will identify these higher order contributions and discuss their effects, focusing on perturbativity considerations when the dynamics exhibit the particle production process described by \eqref{tpk}.  
\section{Non-linearities in the non-Abelian gauge sector} \label{sec3}
\begin{figure}[t!]
\begin{center}
\includegraphics[scale=0.4]{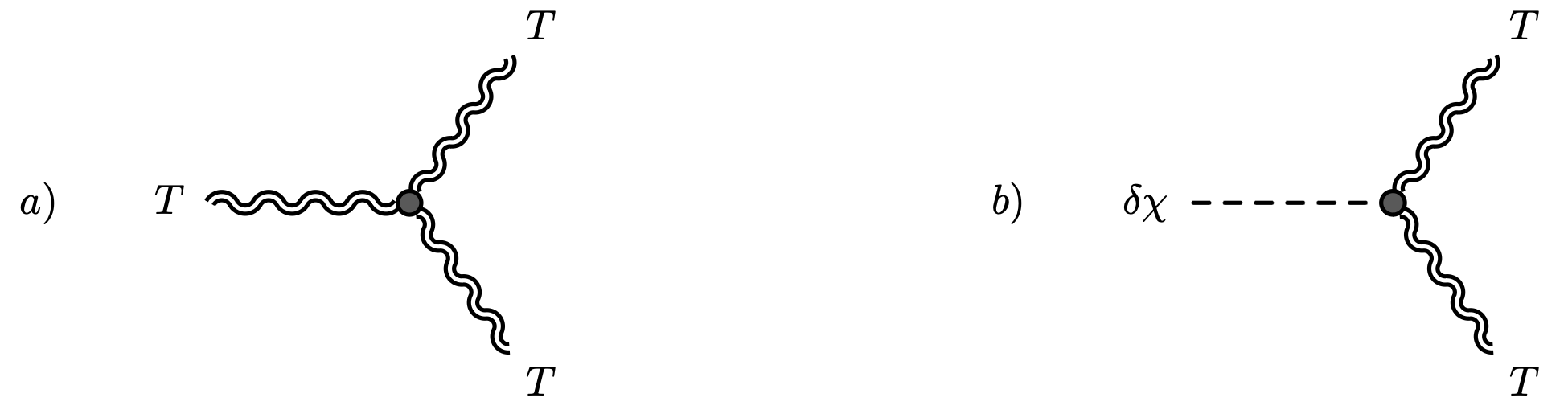}
\end{center}
\vspace*{-7mm}\caption{Three-point vertices of $ttt$ (left) and $tts$ (right) type. The interaction on the left is proportional the gauge coupling $g$ and is universally present for any model endowed with a \emph{non-Abelian} gauge sector. The diagram on the right characterizes the axion gauge field interaction in Eq.~\eqref{Lint}.\label{fig:vertices}}
\end{figure}
In this section, we study the effects of interactions beyond the linear level. In particular, our aim is to derive leading-order corrections --induced at one-loop order-- to the two-point correlator of the $T_+$ mode. For this purpose, we first identify the interactions that are relevant to the loop effects. By expanding the part of the action \eqref{LAGF} that contains gauge field fluctuations, two terms at cubic order in the perturbations are of particular interest:
\begin{enumerate}
    \item The gauge field kinetic term contains interactions of tensor-tensor-tensor ($ttt$) type, due to the non-Abelian nature of the gauge fields, described by ($ S \equiv \int \d^3 x\,\d \tau\, \mathcal{L}$):
    \beq\label{Lttt}
    \mathcal{L}_{ttt} = -g \left[\epsilon^{abc}T_{ai}T_{bj}\left(\partial_iT_{cj}+\frac{1+m_Q^2}{3m_Q\tau}\epsilon^{ijk}T_{ck}\right)+\frac{m_Q}{\tau}T_{ij}T_{jk}T_{ki}\right]\,.
    \eeq
    This Lagrangian was recently employed for the calculation of tensor non-Gaussianities induced by the gauge sector \cite{Agrawal:2017awz,Agrawal:2018mrg}\footnote{The different sign, compared to \cite{Agrawal:2017awz,Agrawal:2018mrg}, in front of the first term is explained by the opposite sign convention chosen for the term proportional to the gauge coupling $g$ in \eqref{gaugedef} and for the Chern-Simons interaction. Such a choice leads to an instability of the opposite chirality which makes up for the sign difference.}. We, on the other hand, are interested in the effect of these interactions on the gauge field modes themselves. We emphasize that this contribution is truly universal in models that exhibit particle production in the gauge sector by the rolling axions, as long as the former has a non-Abelian gauge group. A sketch of the relevant vertex for these interaction is shown in the left panel of Fig. \ref{fig:vertices}
    \item The very same interaction that leads to particle production, namely the Chern-Simons term \eqref{Lint}, contains terms of the tensor-tensor-scalar ($tts$) type, which can potentially induce a large loop correction to the lowest order correlator of the gauge modes. The relevant cubic Lagrangian for this interaction is given by \cite{Papageorgiou:2018rfx,Dimastrogiovanni:2018xnn}
    \beq\label{Ltts}
    \mathcal{L}_{tts} = \frac{\lambda}{f} \left\{\delta\chi\left[\frac{g}{2}(aQ\, T_{ab}T_{ab})'-\epsilon^{ijk}T_{ai}'\partial_jT_{ak}\right]+\left[\frac{g^2a^2Q^2}{-\partial^2+2g^2a^2Q^2}\delta\chi\right]\partial_j\left(\epsilon^{ijk}T_{ai}'T_{ak}\right)\right\},
    \eeq
    where the non-local $\partial^2$ term should be considered as acting on the Fourier modes of the axion perturbation $\delta\chi$. We note that the corresponding correction induced at loop level is model-dependent because of its dependence on the axion perturbations, whose behavior is much more sensitive (than $T$) to different background evolutions. We will explore the $tts$ effect on the gauge mode propagator by focusing on the model recently proposed in \cite{Dimastrogiovanni:2023oid}. The corresponding representative diagram for the $tts$ interaction is shown in the right panel of Fig \ref{fig:vertices}.
\end{enumerate}

One can verify that these interactions are responsible for the dominant contributions to the perturbative correction to $\langle T_+T_+\rangle$. Indeed, a $\delta\chi^2t$ vertex would result in a sub-leading contribution (see e.g. \cite{Dimastrogiovanni:2018xnn,Papageorgiou:2018rfx}), while any other coupling would either include metric perturbations or scalar perturbations in the gauge sector, all of which are typically suppressed with respect to interaction terms containing axion perturbation (see e.g. \cite{Dimastrogiovanni:2012ew,Dimastrogiovanni:2023oid}).
A loop of (enhanced) metric tensors would also be (slow-roll) suppressed relative to a loop of gauge tensors due to the $h-t$ coupling. While an exact result would require to take into account all these additional interactions, the leading-order result can be obtained from \eqref{Lttt} and \eqref{Ltts}. 
\begin{figure}[t!]
\begin{center}
\includegraphics[scale=0.4]{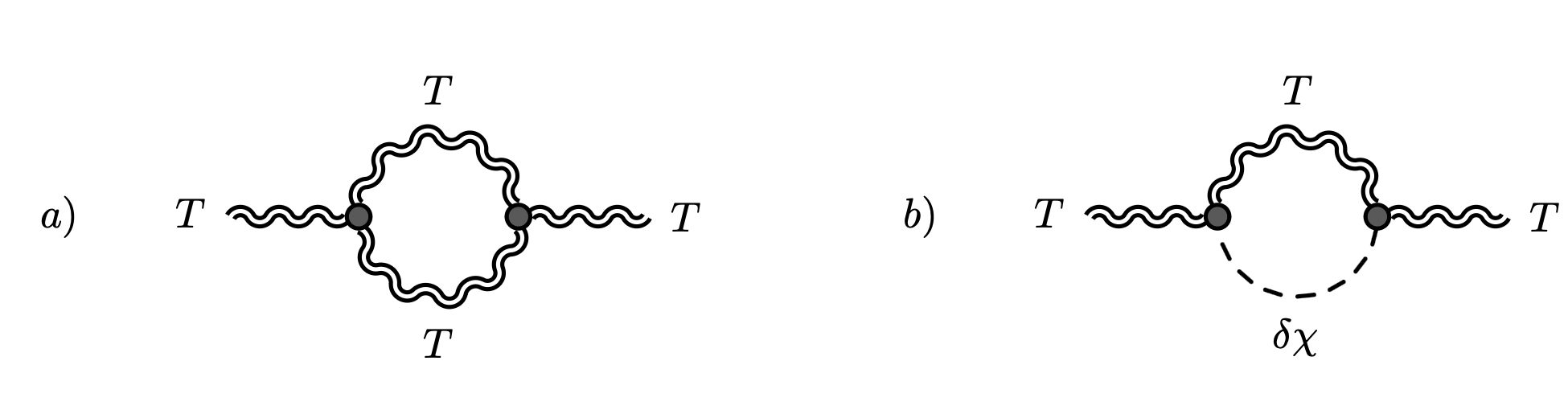}
\end{center}
\vspace*{-10mm}\caption{Feynmann diagrams of the leading-order one-loop corrections to the gauge field propagator by the 3-vertices presented in Fig \ref{fig:vertices}. The corresponding contributions are presented in Eqs.~\eqref{Pttt}-\eqref{P2ttt} and \eqref{P1tts}, respectively. The diagram on the left ($\propto g^2$) is universally present in models with non-Abelian gauge fields.\label{fig:oneloop}}
\end{figure}

In our models of interest, a large gauge field fluctuations can source cosmological perturbations, such as chiral GWs, at observable levels, which makes these models phenomenologically interesting. In these cases one may wonder whether the
large amplitudes of the gauge field fluctuations can drive the system out of the
perturbative regime. In the following analysis, our aim is to identify the regime of perturbative control. For this purpose, we focus on the tensor degrees of freedom in the gauge sector and require that higher-order loop corrections induced through the interactions \eqref{Lttt} and \eqref{Ltts} do not spoil the leading-order (tree level) estimates for the amplified gauge field mode functions. This criterion can be written more precisely as
\beq\label{pertc}
\mathcal{R}^{(\alpha)}_T\left(m_Q,\tau\right) \equiv\left|\frac{\delta^{(1)}_{(\alpha)}\left\langle \hat{T}_{+}(\tau, \vec{k})\, \hat{T}_{+}(\tau, \vec{k}^{\prime})\right\rangle^{\prime}}{\left\langle \hat{T}_{+}(\tau, \vec{k}) \, \hat{T}_{+}(\tau, \vec{k}^{\prime})\right\rangle_{\rm tree}^{\prime}}\right| \ll 1,
\eeq
where we defined the one-loop correction in the numerator with the $\delta^{(1)}\langle\dots\rangle$ notation, $\alpha = \{ttt,tts\}$ labels the vertex under consideration (see Fig. \ref{fig:vertices}) and the prime denotes the two-point function without the corresponding delta function. The schematic one-loop diagrams corresponding to these interactions are shown in Fig.~\ref{fig:oneloop}.

Next, we discuss briefly the computation of the relevant one-loop expectation values. 
From now on we restrict our attention to the enhanced $T_+$ mode that is amplified in the slow-roll trajectory of the axion-gauge field system.
We define the one-loop power spectrum $P^{(1)}_{\alpha}$ of the gauge field as 
\beq\label{dPS}
\Big\langle\, \hat{T}_{+}(\tau,\vec{k})\, \hat{T}_{+}(\tau,\vec{k}') \,\Big\rangle_{\rm 1-loop} \equiv \delta^{(1)}_{(\alpha)}\Big\langle\, \hat{T}_{+}(\tau,\vec{k})\, \hat{T}_{+}(\tau,\vec{k}') \,\Big\rangle  = \delta^{(3)}\left(\vec{k}+\vec{k'}\right) P^{(1)}_{\alpha}(\tau, k),
\eeq
where the gauge tensor perturbations are promoted to quantum operators as
\beq\label{that}
\hat{T}_{\lambda} (\tau,\vec{k}) =  T_{\lambda}(\tau,k)\, \hat{a}_\lambda(\vec{k}) + T_{\lambda}(\tau,k)^{*} \hat{a}_\lambda^{\dagger} (-\vec{k}),
\eeq
with the creation and annihilation operators satisfying the canonical commutation relation
\beq
\left[\hat{a}_\lambda(\vec{k}) ,\hat{a}_\lambda'^{\dagger} (-\vec{k}')\right]=\delta^{\lambda\lambda'}\delta^{(3)}(\vec{k}+\vec{k}').
\eeq

One can compute one-loop power spectra with the {\it in-in formalism} \cite{Weinberg:2005vy,Adshead:2009cb} as
\beq\label{TTin}
    \delta^{(1)}_{(\alpha)}\Big\langle\, \hat{T}_{+}(\tau,\vec{k})\, \hat{T}_{+}(\tau,\vec{k}') \,\Big\rangle=-\int^{\tau}_{-\infty}\d\tau'\int^{\tau'}_{-\infty}\d\tau''\,\Big\langle\left[\left[\hat{T}_{+}(\tau,\vec{k})\, \hat{T}_{+}(\tau,\vec{k}'),\hat{H}_{\alpha}^{(3)}(\tau')\right],\hat{H}_{\alpha}^{(3)}(\tau'')\right]\Big\rangle,
\eeq
where  ($\hat{H} \equiv -\int \d^3 x\, \mathcal{\hat{L}}$) is the interaction Hamiltonian, consisting of fields that are evolved by the free Hamiltonian, as usual in the interaction picture. The perturbativity criterion we defined in Eq.~\eqref{pertc} can then be simply re-written as
\beq\label{pertcf}
    \mathcal{R}^{(\alpha)}_{T}(m_Q, \tau)\equiv \bigg|\frac{P^{(1)}_{\alpha}(\tau,k)}{P^{(0)}(\tau,k)}\bigg|\ll 1\,, 
\eeq
where the tree-level power spectrum is defined by the following two-point function
\beq\label{dPV}
  \Big\langle\hat{T}_+(\tau,\vec{k}) \hat{T}_+(\tau,\vec{k}')\Big\rangle_{\rm tree} =  \delta^{(3)}\left(\vec{k}+\vec{k'}\right) P^{(0)}(\tau, k),
\eeq
with
\beq\label{rrr}
P^{(0)}(\tau, k) \equiv \Big|T_+(\tau,k)\Big|^2=\frac{{\rm e}^{\pi(2m_Q+m_Q^{-1})}}{2k}\,\Big|W_{\alpha,\beta}(-2ix)\Big|^2\,,
\eeq
where we used the analytical solution introduced in Eq. \eqref{tpk}. On general grounds, we expect the loop contribution coming from the \emph{tts} vertex to be smaller than the \emph{ttt} counterpart, as the former includes fewer factors of $T_+ \propto {\rm e}^{\pi(2 m_Q + m_Q^{-1})/2}$. We will confirm this expectation by focusing on the specific setup recently proposed in \cite{Dimastrogiovanni:2023oid}. In the following subsections, we derive and discuss the \emph{ttt} and \emph{tts} contributions separately. 

\subsection{One-loop spectra from the self-interactions of the gauge field}
The one-loop power spectrum of the gauge field tensor modes induced by the self interactions can be computed with Eq.~\eqref{TTin} and the following interaction Hamiltonian

\beq\label{Httt}
    \hat{H}_{ttt}^{(3)}(\tau)=g \int\,\d^3x\,\left[\epsilon^{abc}\hat{T}_{ai}\hat{T}_{bj}\left(\partial_i\hat{T}_{cj}+\frac{1+m_Q^2}{3m_Q\tau}\epsilon^{ijk}\hat{T}_{ck}\right)+\frac{m_Q}{\tau}\hat{T}_{ij}\hat{T}_{jk}\hat{T}_{ki}\right].
\eeq
Note that the perturbations entering the right hand side of \eqref{TTin} and \eqref{Httt} are those obtained in the linear theory presented in the previous section.

Using the decomposition of the gauge tensor field $\hat{T}_{ab}$ \eqref{tft}-\eqref{Piab} in the Hamiltonian \eqref{Httt}, we carried out the expectation value of the nested commutators in \eqref{TTin}, the details of which can be found in Appendix \ref{AppA}. Recalling the definition in Eq.~\eqref{dPS}, one finds that the one-loop power spectrum induced by the self interactions of the gauge field is made up by two distinct contributions:
\beq\label{Pttt}
P^{(1)}_{ttt}(\tau, k) = \sum_{i = 1}^{2} P^{(1,i)}_{ttt}(\tau, k).
\eeq
The first contribution is the standard expression one would have obtained employing a Green's function approach with the sources bilinear in the gauge fields:  
\beq\label{P1ttt}
    P_{ttt}^{(1,1)}(\tau, k)=\fr{g^2}{4\pi^3}\int\,\d^3q
    \left|\int^{\tau}_{-\infty}\d\tau'\left[\epsilon_1+\frac{1+m_Q^2}{m_Q\tau'}\epsilon_2+\frac{3m_Q}{\tau'}\epsilon_3\right]\mathcal{G}^{(k)}_T(\tau,\tau')T_+(\tau',q)T_+(\tau',|\vec{k}-\vec{q}|)\right|^2,
\eeq
where the products of polarization vectors $\epsilon_i \equiv \epsilon_i (-\vec{k},\vec{q},\vec{k}-\vec{q})$ ($i = 1,2,3$), invariant under the exchange of any of the arguments, are reported in \eqref{ept}. We note that the Green's function that appears inside the integral is given by \cite{Agrawal:2018mrg} 
\begin{align}\label{GreenT}
    \nn \mathcal{G}^{(k)}_T(\tau,\tau') &= i\theta(\tau-\tau')\left[T_+(\tau,k)T^{*}_+(\tau',k)-T^{*}_+(\tau,k)T_+(\tau',k)\right]\\ 
    &=\frac{1}{k}\,\theta(\tau-\tau')\, {\rm e}^{\pi (2m_Q + m_Q^{-1})}\, {\rm Im}\left[ W_{\alpha,\beta}^{*}(2ik\tau)\, W_{\alpha,\beta}(2ik\tau')\right].
\end{align}
The second contribution is given by
\begin{align}\label{P2ttt}
    \nn P^{(1,2)}_{ttt}(\tau,k)&=\frac{2g^2}{\pi^3} \int\d^3q\int^{\tau}_{-\infty}\d\tau'\int^{\tau'}_{-\infty}\d\tau''\left[\epsilon_1+\frac{1+m_Q^2}{m_Q\tau'}\epsilon_2+\frac{3m_Q}{\tau'}\epsilon_3\right] \left[\epsilon_1+\frac{1+m_Q^2}{m_Q\tau''}\epsilon_2+\frac{3m_Q}{\tau''}\epsilon_3\right]^* \\ \nn
    &\quad\quad\quad\quad\quad\quad\quad\quad\quad\quad \times{\rm Im}\left[T_+(\tau,k)T^*_+(\tau',k)\right]\,{\rm Re}\left[T_+(\tau,k)T^*_+(\tau'',k)\right]\\
    &\quad\quad\quad\quad\quad\quad\quad\quad\quad\quad \times \,{\rm Im}\left[T_+(\tau', q)T^*_+(\tau'',q)T_{+}(\tau', |\vec{k}-\vec{q}|)T^*_{+}(\tau'', |\vec{k}-\vec{q}|)\right].
\end{align}

\subsection{One-loop spectra through interactions of \emph{tts} type} 
The perturbative corrections induced by the tensor-tensor-scalar interaction \eqref{Ltts} can be obtained in an analogous way, by plugging into Eq.~\eqref{TTin} the following interaction Hamiltonian
\beq \label{Htts}
        \hat{H}_{tts}^{(3)}(\tau) =-\fr{\lambda}{f}\int\,\d^3x\,\left\{\hat{\delta\chi}\left[\frac{g}{2}(aQ\hat{T}_{ab}\hat{T}_{ab})'-\epsilon^{ijk}\hat{T}_{ai}'\partial_j\hat{T}_{ak}\right]+\left[\frac{g^2a^2Q^2}{-\partial^2+2g^2a^2Q^2}\hat{\delta\chi}\right]\partial_j\left(\epsilon^{ijk}\hat{T}_{ai}'\hat{T}_{ak}\right)\right\}\,. \\
\eeq
We provide the details of the calculation in Appendix \ref{AppA} and report here the final result:
\beq \label{P1tts}
\begin{aligned}
P^{(1)}_{tts}(\tau, k) &=\frac{\lambda^2H^2}{\pi^3f^2F_g}\int\,\d^3q\,\left|\vec{\epsilon}(-\hat{k})\cdot\vec{\epsilon}(\hat{q})\right|^4\int^{\tau}_{-\infty}\d\tau'\int^{\tau'}_{-\infty}\d\tau'' \mathcal{D}_{\tau_1'\tau_2'}\,\mathcal{D}_{\tau_1''\tau_2''}\,{\rm Im}\left[T_+(\tau,k)T^*_+(\tau_1',k)\right] \\
&\quad\quad\quad\times {\rm Im}\left[T_+(\tau,k)T^*_+(\tau_1'',k)T_+(\tau_2', q)T^*_+(\tau_2'',q)X(\tau',q')X^*(\tau'',q')\right]\Big|_{\tau_1'=\tau_2'=\tau',\tau_1''=\tau_2''=\tau''},
\end{aligned}
\eeq
where the operator $\mathcal{D}$ acting on the gauge tensors is defined as
\beq
\mathcal{D}_{\tau_1\tau_2}\equiv \frac{m_Q}{\tau} -\left[\frac{m_Q}{\tau}+q+\frac{(k-q)\,m_Q^2}{|\vec{k}-\vec{q}|^2\tau^2+2m_Q^2}\right]\tau\partial_{\tau_1}-\left[\frac{m_Q}{\tau}+k+\frac{(q-k)\,m_Q^2}{|\vec{k}-\vec{q}|^2\tau^2+2m_Q^2}\right]\tau\partial_{\tau_2}\,.
\eeq
Notice that the one-loop result \eqref{P1tts} contains the mode function of the axion, which is typically model-dependent as we discussed in section \ref{sec2}. This makes the corresponding one-loop contribution also model-dependent, in spite of the three-point interaction vertex \eqref{Ltts} being common to all possible scenarios endowed with a Chern-Simons coupling \eqref{Lint}. In the next section, we present our results for this diagram focusing on a particular setup that includes a non-minimal coupling with the axion-inflaton \cite{Dimastrogiovanni:2023oid}. A review of the relevant dynamics of the axion perturbation in this scenario can be found in Appendix \ref{AppB}. Crucially, in this setup the axion perturbation behaves as a free massless scalar field, for which the analytical solution is known (Eq.~\eqref{Xc}). \\
\indent In the next section, we explicitly compute the perturbativity conditions \eqref{pertcf} arising from \eqref{P1ttt}, \eqref{P2ttt} and \eqref{P1tts}. 

\subsection{Perturbativity of the gauge field fluctuations}
Following the explicit steps detailed in Appendix \ref{AppA}, we recast the perturbativity conditions \eqref{pertcf} for the two interactions using $x = -k\tau$ as the external time variable, and evaluate them at $x = x_{\rm p}$ to obtain 
\begin{align}\label{Rs}
\nn 
\mathcal{R}^{(ttt)}_T (x_{\rm p}, m_Q) &= \fr{g^2}{2\pi^2}\, f_{2,ttt}(x_{\rm p}, m_Q)\ll1\,,\\
\mathcal{R}^{(tts)}_T (x_{\rm p}, m_Q) &=  \frac{\lambda^2H^2}{64  \pi^2f^2F_g} f_{2,tts}(x_{\rm p},m_Q)\ll1\,.
\end{align}
We introduced here the normalized functions $f_{2,\alpha}$ (Eqs.~\eqref{f2tttapp}-\eqref{f2tttapp2} and \eqref{f2ttsapp}) parametrizing the loop factors induced by the non-linearities. 

We note that $F_g$ in \eqref{Rs} is a factor associated with our choice of a non-minimally coupled axion inflation scenario \cite{Dimastrogiovanni:2023oid}. As we discuss in Appendix \ref{AppB}, $F_g$ is a canonical normalization factor for the axion perturbations. The expressions above (in combination with \eqref{f2ttsapp}) can be applied to any other model, after choosing $F_g$ appropriately, as long as the time variation of $F_g$ is weak, \ie $\dot{F_g}/(H F_g) \ll 1$. As an example, for a canonical axion sector, one would simply set $F_g = 1$.

\begin{figure}[t!]
\begin{center}
\includegraphics[scale=0.464]{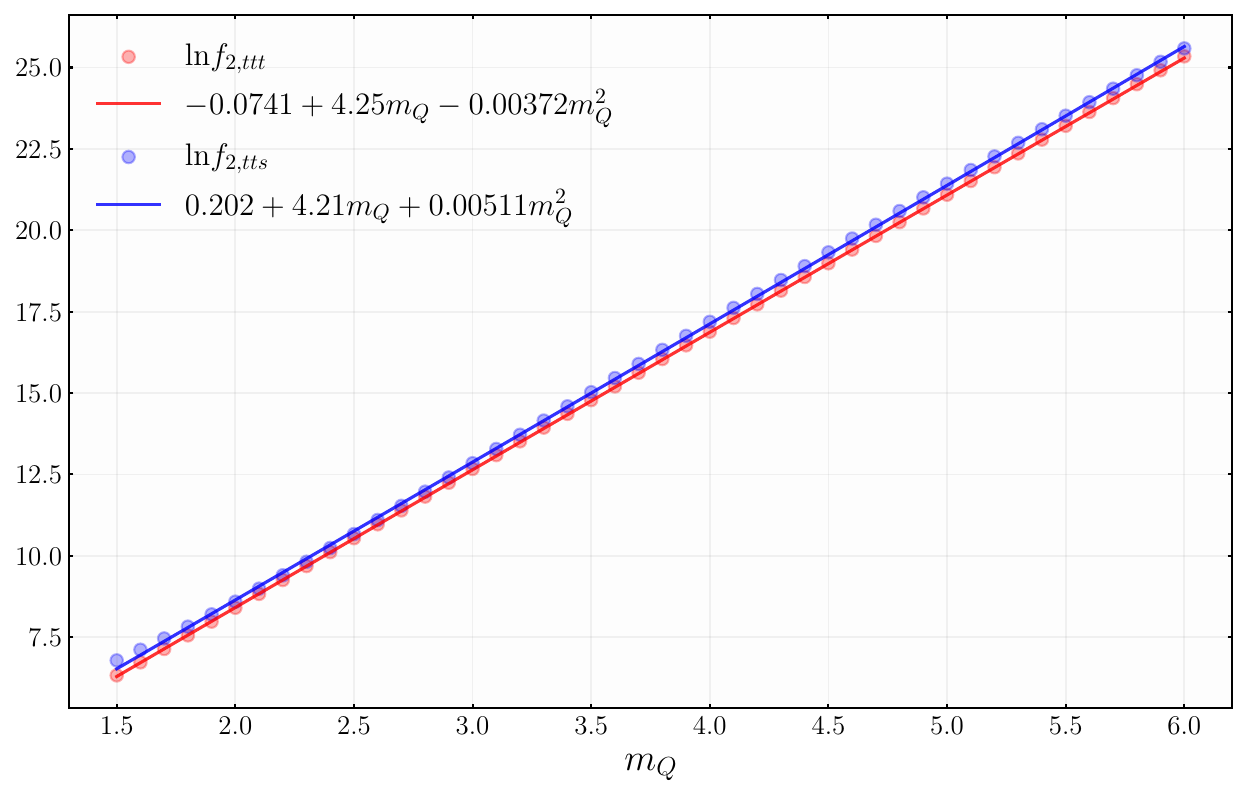}
\end{center}
\vspace*{-5mm}\caption{Normalized loop corrections ${\rm ln}f_{2,\alpha}(x_{\rm p},m_Q)$ as a function of $m_Q$ (dots) for both diagrams $\alpha = \{ttt,tts\}$ we consider, together with the accurate fitting forms \eqref{f2s} (solid lines) that parametrize the numerical results in terms of $m_Q$, characterizing the strength of particle production in the gauge sector. \label{fig:f2}}
\end{figure}
It is clear from Eq. \eqref{Rs} that the perturbativity criterion has an explicit $m_Q$ dependence, mainly reflecting the influence of particle production processes in the gauge sector. 
Implementing a cut-off procedure as explained in Appendix \ref{AppA}, we  numerically evaluate the normalized loop corrections ${f}_{2,\alpha}$ at the time corresponding to the gauge field sources peak (see Figure \ref{fig:Tp} and Eq. \eqref{xp}). As mentioned earlier, the solution \eqref{tpk} we use in the loop computations neglects the mixing with the metric tensor perturbations (relevant only outside the horizon for $x\ll 1$ \cite{Adshead:2012kp}). 
Consistently with this choice, we evaluate the normalized loop factors $f_{2,\alpha}$ at $x = x_{\rm p}$ where the gauge tensor modes are maximally enhanced.\\
\indent Computing the normalized loop factors (see Eqs.~\eqref{f2tttapp},\eqref{f2tttapp1},\eqref{f2tttapp2} and \eqref{f2ttsapp}) for a grid of $m_Q$ values within $1.5\leq m_Q\leq6 $, we obtain the following fitting forms for $f_{2,\alpha}$ in terms of the enhancement parameter $m_Q$: 
\begin{align}\label{f2s}
\nn \ln\left[f_{2,ttt}(x_{\rm p},m_Q)\right]&\simeq {-0.0741+4.25m_Q-0.00372m_Q^2}, \\
\ln\left[f_{2,tts}(x_{\rm p},m_Q)\right]&\simeq {0.202+4.21m_Q+0.00511m_Q^2}.
\end{align}
The accuracy of these fitting functions with respect to the numerical results are shown in Fig.~\ref{fig:f2}. As expected, the perturbativity criterion becomes more demanding as the value of $m_Q$ increases.

\subsection{Implications of perturbativity constraints on axion gauge field models}
Armed with the results we derived in the previous section, we now turn to discussing the perturbativity criteria \eqref{Rs} taking into account the loop pre-factors of the corresponding diagrams.  
\medskip

\begin{figure}[t!]
\begin{center}
\includegraphics[scale=0.464]{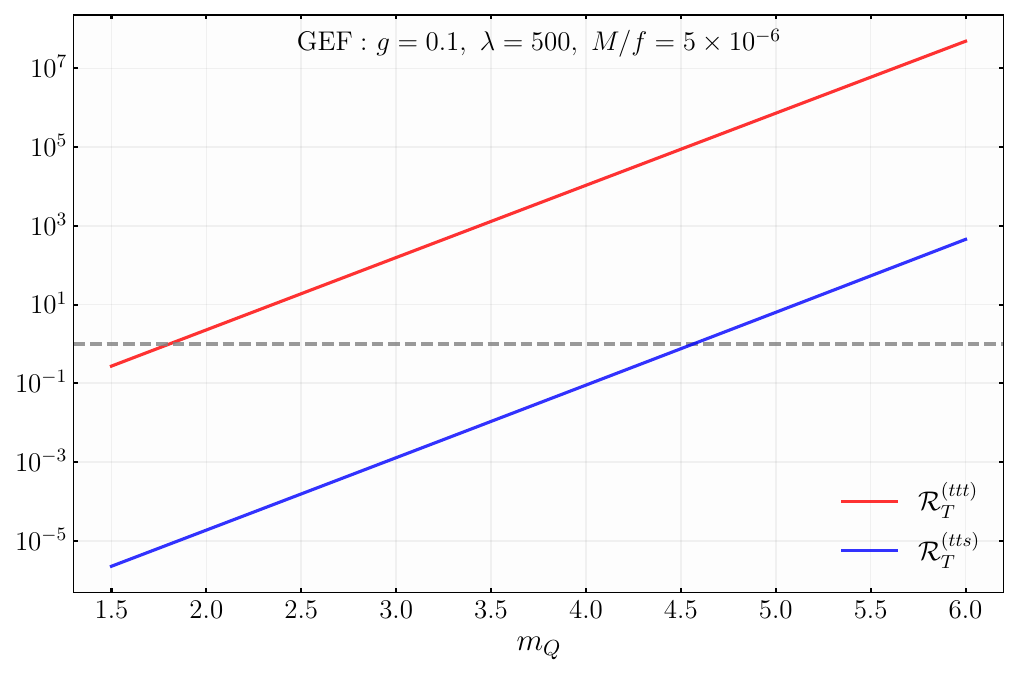}\includegraphics[scale=0.464]{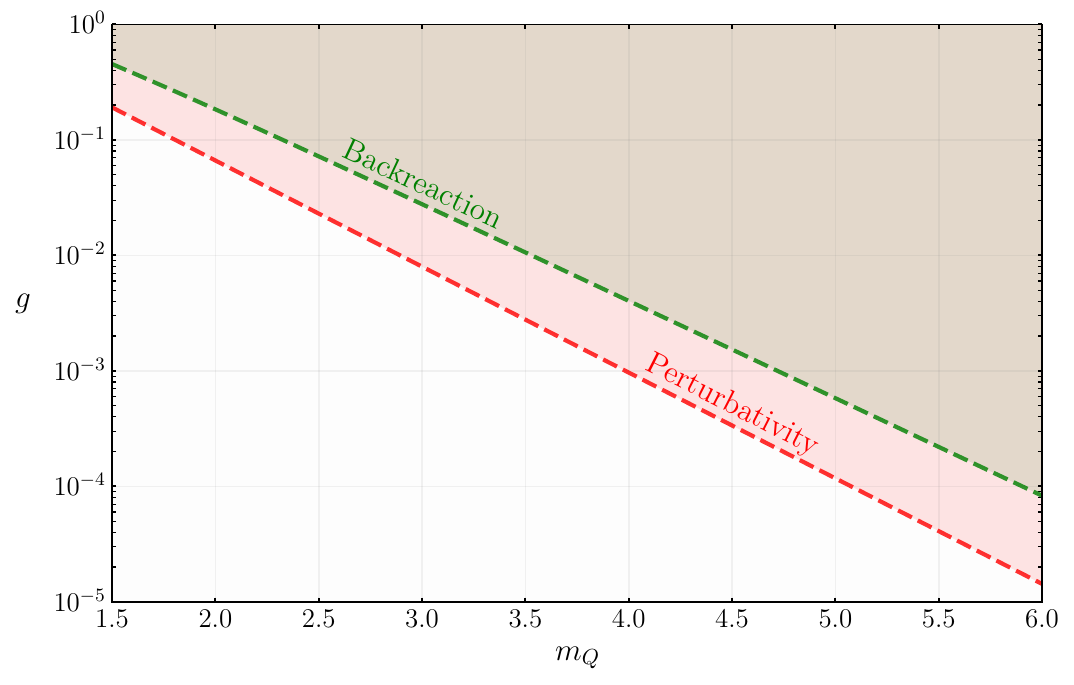}
\end{center}
\vspace*{-5mm}\caption{The perturbativity criteria in Eqs.~\eqref{Rtts} and \eqref{Rttt} in the non-minimally coupled axion-gauge field model of \cite{Dimastrogiovanni:2023oid}, for fiducial values of the parameters $g$, $\lambda$ and $M$ (left). The universal perturbativity constraint \eqref{Rttt}, as compared to the backreaction constraint \eqref{tbr}, in the 
 $m_Q$ vs $g$ parameter space (right). \label{fig:mqvsg}}
\end{figure}

\noindent{\bf \textsl{tts} contribution to perturbativity}.  The perturbativity condition for the one-loop diagram with the tensor-tensor-scalar vertex can be written as
\beq\label{Rtts}
\mathcal{R}^{(tts)}_T (x_{\rm p}, m_Q) =  \frac{\lambda^2H^2}{64  \pi^2f^2F_g}\,\, {\rm e}^{{0.202+4.21m_Q+0.00511m_Q^2}} \ll 1\; ,
\eeq
where we simply combined Eqs.~\eqref{Rs} and \eqref{f2s}. As we already pointed out, this result is not universal among all the inflationary models with the interaction \eqref{LAGF}. This is because the corresponding loop diagram is sensitive to the dynamics of the axion sector, whose behavior in general depends on the details of the slow-roll trajectory at the background level. In particular, the result we present above is valid for the inflationary scenario studied in \cite{Dimastrogiovanni:2023oid}, characterised by a non-minimal coupling of the axion inflaton to the gravitational sector. The $m_Q$-dependent fitting form in the exponential argument is derived focusing on the behavior of the axion perturbations, which in this specific scenario essentially acts like a free scalar field in a (quasi) de-Sitter background. More details on these results can be found in Appendix \ref{AppB}.  In the particular setup we are working on, there is an extra source of suppression originating from the loop pre-factors in \eqref{Rtts} which satisfy  
\beq
\frac{\lambda^2 H^2}{64\pi^2f^2 F_g} \equiv \frac{g^2}{32\pi^2\,m_Q^2}\frac{\Lambda^2}{2F_g}\ll \frac{g^2}{2\pi^2}\,,
\eeq
from having $\Lambda^2 \ll 2F_g$ in the so called gravitational enhanced friction (GEF) regime \cite{Dimastrogiovanni:2023oid}. The left panel of Figure \ref{fig:mqvsg} illustrates the hierarchy $\mathcal{R}^{(ttt)}_T\gg \mathcal{R}^{(tts)}_T$ for all $m_Q$ values of phenomenological interest during inflation. 
\medskip

\noindent{\bf A universal constraint}.  We now turn to discussing the universal constraint arising through the self-interactions of the gauge field. The perturbativity criterion for the one-loop diagram containing the tensor-tensor-tensor vertex reads 
\beq\label{Rttt}
\mathcal{R}^{(ttt)}_T (x_{\rm p}, m_Q) = \fr{g^2}{2\pi^2}\, {\rm e}^{-0.074+4.25m_Q-0.00372m_Q^2}\ll 1\; ,
\eeq
where what we call its universal nature rests upon it being simply dependent on the gauge coupling $g$ and the parameter $m_Q$ that quantifies the amplification of the gauge mode functions around horizon crossing. We emphasize that Eq.~\eqref{Rttt} holds whenever the inflationary background dynamics is in the slow-roll attractor regime described by Eqs.~\eqref{Qmin} and \eqref{theR}, irrespective of the details regarding the axion sector, including its passive/active role in driving the accelerated expansion. 

In the left panel of Fig.~\ref{fig:mqvsg}, we plot Eq.~\eqref{Rttt} (red line) and Eq.~\eqref{Rtts} (blue line) as a function of $m_Q$. We find that, for a gauge coupling as large as $g = 0.1$\footnote{One could lower the value of g while remaining within the GEF regime, however the agreement of the model predictions with CMB observations would in that case worsen, as shown in \cite{Dimastrogiovanni:2023oid}.}, the perturbativity condition that originates from the gauge field self-interactions is very demanding already at the initial stages of inflation ($m_Q \sim  2$).

In the right panel of Fig.~\ref{fig:mqvsg}, we show the $\mathcal{R}^{(ttt)}_T < 1$ region in the $m_Q-g$ parameter space alongside the semi-analytical estimates on the  backreaction constraint\footnote{This is the backreaction of the gauge field on the dynamics of its vev $Q$.}\cite{Dimastrogiovanni:2016fuu,Papageorgiou:2019ecb}:
\beq\label{tbr}
g\ll\left(\frac{24\pi^2}{\mathcal{B}-\tilde{\mathcal{B}}/\xi}\frac{1}{1+1/m_Q^2}\right)^{1/2}\,.
\eeq
Here $\left(\mathcal{B}-\tilde{\mathcal{B}}/\xi\right)\simeq2.3{\rm e}^{3.9m_Q}$.
We see from the Figure that \eqref{Rttt} gives rise to a stronger condition than \eqref{tbr} (green region). The allowed parameter space corresponds to the remaining white region. These results imply that the perturbativity of the gauge sector imposes a condition at least as strong as the \textsl{naive} backreaction constraints\footnote{We note that similar conditions arise for the ${\rm U}(1)$ cousin(s) of axion-gauge field models we consider, see \eg \cite{Peloso:2016gqs, Ozsoy:2020ccy, Campeti:2022acx} for studies pointing in this direction.} stemming from requiring one stays clear from the strong backreaction regime, as typically done in the literature. From a dynamical perspective these results point to the fact that any given axion-gauge field model, initialized in the safe zone (white region), would become non-perturbative (moving horizontally along the increasing $m_Q$ direction in Fig.~\ref{fig:mqvsg}) before entering the strong backreaction regime.

\section{Conclusions} \label{sec4}

Inflationary models equipped with a Chern-Simons coupling between axions  and gauge fields have an interesting phenomenology: large gauge field perturbations, sustained by the energy of the rolling axion, can be generated and in turn source scalar and tensor fluctuations leading to potentially observable signatures. However, the copious production of gauge quanta comes with two additional effects: (i) gauge field fluctuations can backreact on the background evolution and (ii) gauge (self)interactions can lead to large loop corrections to cosmological propagators. When the effects in (i) become important, the system enters the so-calxled strong backreaction regime, one that requires dedicated studies \cite{Caravano:2021bfn,Figueroa:2023oxc,Iarygina:2023mtj,Domcke:2023tnn,Dimastrogiovanni:2024xvc} accounting for the modification of the evolution equations. With regard to point (ii), when loop contributions to the gauge field propagator consistently exceed its tree-level value, the theory is no longer under perturbative control.\\
\indent In this work, we tackled the issue of perturbative control in non-Abelian gauge field scenarios by computing the mode function renormalization of the gauge field. We identified the leading-order contribution arising from the self-interaction (\textsl{ttt}-type) of the growing tensor mode ``$t^{}_{\rm ij}$''  in the gauge sector, finding
that is largely insensitive to the details of the axion Lagrangian (including, e.g., the nature of the axion couplings to gravity, its specific potential, as well as its role as inflaton or spectator field). The value of the \emph{ttt}-generated loop correction is \emph{universal}, the only condition in place being that the background be moving on a slow-roll trajectory within the weak backreaction regime. We also computed the contribution to the  $t$-propagator arising from a \emph{tts}-interaction between the gauge field and the axion. Unlike the \emph{ttt}-sourced loop, the \emph{tts}-one is rather model-dependent and so we calculated its value in a specific scenario \cite{Dimastrogiovanni:2023oid}. 

We then proceeded to compare the limits on the parameter space arising from perturbativity constraints with those associated with the onset of backreaction, finding that the perturbativity constraints are more stringent. Our results are summarised in Fig.~\ref{fig:mqvsg}. Dynamically, our findings imply that a growing value of the particle production parameter $m_Q$ (or the related $\xi$ parameter) during inflation would cause the system to reach the limit set by perturbativity before entering the strong backreaction region.  \\
\indent Our analysis could be further developed in a number of directions. First of all, our results for the perturbativity constraints were obtained semi-analytically, neglecting the corrections from the mixing of the tensor modes in the gauge sector with the metric. Metric fluctuations are indeed  known to have a slight effect on  $t$ modes when these are sufficiently outside the horizon. In this regard, the perturbativity limits we obtained are slightly underestimated. It would be interesting to perform the fully numerical computation so as to include also the super-horizon correction. In our analysis we also disregarded a number of sub-leading diagrams, which could be included to deliver a more accurate (and possibly slightly more) stringent perturbativity bound. We should also stress that the parameter space corresponding to the onset of strong backreaction was found using the (analytical) estimates present in the literature. A more sophisticated analysis involving numerical techniques and/or lattice simulations is a natural next step in the direction of deriving the full extent of the interplay between backreaction effects and perturbativity bounds. We leave this to future work.

\acknowledgments
We are indebted to Alexandros Papageorgiou for useful discussions. We thank the Center for Information Technology of the University of Groningen for their support and for providing access to the Hábrók high performance computing cluster. The work of MF and O\"O is partially supported by the Spanish Research Agency (Agencia
Estatal de Investigaci\'on) through the Grant IFT Centro de Excelencia Severo Ochoa No CEX2020-001007-S,
funded by MCIN/AEI/10.13039/501100011033. MF acknowledges the ``Consolidaci\'on Investigadora'' grant CNS2022-135590 and from the ``Ram\'on y Cajal'' grant RYC2021-033786-I.  O\"O is supported by the “Juan de la Cierva” fellowship IJC2020-045803-I. O\"O would also like to thank Perimeter Institute for Theoretical Physics and especially Niayesh Afshordi for hospitality while some parts of this work were being completed. 

\newpage
\begin{appendix}

\section{Non-linear sources of gauge field perturbations} \label{AppA}
\subsection*{Gauge field renormalization by the \emph{ttt} interaction}
We provide here some details of the in-in calculations that leads to the result \eqref{P1ttt} in the main text. We begin by performing a Fourier mode decomposition \eqref{tft} for $\hat{T}_{ab}$ in the interaction Hamiltonian \eqref{Httt} 
\begin{align}\label{Hft}
    \hat{H}_{ttt}^{(3)}(\tau)&=g \int\,\frac{\d^3q_1\d^3q_2\d^3q_3}{(2\pi)^{3/2}}\,\delta(\vec{q}_1+\vec{q}_2+\vec{q}_3)\left[\fr{\epsilon_1}{3}+\frac{1+m_Q^2}{3m_Q\tau}\epsilon_2+\frac{m_Q}{\tau}\epsilon_3\right]_{\vec{q}_1,\vec{q}_2,\vec{q}_3}\\ \nn
    &\quad\quad\quad\quad\quad\quad\quad\quad\quad\quad\quad\quad\quad\quad\quad\quad\quad\quad\quad\quad\quad\quad\quad\quad\times\hat{T}_+(\tau,\vec{q}_1)\,\hat{T}_+(\tau,\vec{q}_2)\,\hat{T}_+(\tau,\vec{q}_3),
\end{align}
where the products of polarization vectors $\epsilon_1, \epsilon_2$, $\epsilon_3$ are given by:
\begin{align}\label{ept}
\nn
\epsilon_1(\vec{q}_1,\vec{q}_2,\vec{q}_3) &= \fr{1}{2}\vec{\epsilon}_+(\vec{q}_1)\cdot\left(\vec{\epsilon}_+(\vec{q}_2)\times\vec{\epsilon}_+(\vec{q}_3)\right) \bigg\{[i(\vec{q}_3-\vec{q}_2)\cdot\vec{\epsilon}_+(\vec{q}_1)]\,[\vec{\epsilon}_+(\vec{q}_2)\cdot\vec{\epsilon}_+(\vec{q}_3)] \\ \nn
&\quad\quad\quad\quad\quad\quad\quad\quad\quad\quad\quad\quad\quad+[i(\vec{q}_2-\vec{q}_1)\cdot\vec{\epsilon}_+(\vec{q}_3)]\, [\vec{\epsilon}_+(\vec{q}_1)\cdot\vec{\epsilon}_+(\vec{q}_2)] \\ \nn
&\quad\quad\quad\quad\quad\quad\quad\quad\quad\quad\quad\quad\quad + [i(\vec{q}_1-\vec{q}_3)\cdot\vec{\epsilon}_+(\vec{q}_2)]\, [\vec{\epsilon}_+(\vec{q}_3)\cdot\vec{\epsilon}_+(\vec{q}_1)] \bigg\}, \\ \nn
     \epsilon_2(\vec{q}_1,\vec{q}_2,\vec{q}_3) &=\left[\vec{\epsilon}_+(\vec{q}_1)\cdot\left(\vec{\epsilon}_+(\vec{q}_2)\times\vec{\epsilon}_+(\vec{q}_3)\right)\right]^2, \\
    \epsilon_3(\vec{q}_1,\vec{q}_2,\vec{q}_3) &=\left[\vec{\epsilon}_+(\vec{q}_1)\cdot\vec{\epsilon}_+(\vec{q}_2)\right]\left[\vec{\epsilon}_+(\vec{q}_2)\cdot\vec{\epsilon}_+(\vec{q}_3)\right]\left[\vec{\epsilon}_+(\vec{q}_3)\cdot\vec{\epsilon}_+(\vec{q}_1)\right], 
\end{align}
and are symmetric under the exchange of any internal momenta.
Adopting a shorthand notation for the $+$ polarization of the gauge field tensors, $\hat{T}_+(\tau,\vec{q})\equiv \hat{T}_{\vec{q}}(\tau)$, we plug the Hamiltonian \eqref{Hft} in the in-in expression \eqref{TTin} to obtain
\begin{align} \label{deltaTT}
   \nn \delta^{(1)}_{(ttt)}\Big\langle\hat{T}_{\vec{k}}(\tau)\, \hat{T}_{\vec{k}'}(\tau)\Big\rangle &=-g^2 \int\, \prod_{i=1}^{6} \frac{\d^3q_i}{(2\pi)^{3}}\,\delta(\vec{q}_1+\vec{q}_2+\vec{q}_3)\,\delta(\vec{q}_4+\vec{q}_5+\vec{q}_6) \int^{\tau}\d\tau'\int^{\tau'}\d\tau'' \\ \nn
    &\times\left[\frac{\epsilon_1}{3}+\frac{1+m_Q^2}{3m_Q\tau'}\epsilon_2+\frac{m_Q}{\tau'}\epsilon_3\right]_{\vec{q}_1,\vec{q}_2,\vec{q}_3}\left[\frac{\epsilon_1}{3}+\frac{1+m_Q^2}{3m_Q\tau''}\epsilon_2+\frac{m_Q}{\tau''}\epsilon_3\right]_{\vec{q}_4,\vec{q}_5,\vec{q}_6} \\ 
    &\times\Big\langle \Big[\left[\hat{T}_{\vec{k}}(\tau)\hat{T}_{\vec{k}'}(\tau),\hat{T}_{\vec{q}_1}(\tau')\hat{T}_{\vec{q}_2}(\tau')\hat{T}_{\vec{q}_3}(\tau')\right],\hat{T}_{\vec{q}_4}(\tau'')\hat{T}_{\vec{q}_5}(\tau'')\hat{T}_{\vec{q}_6}(\tau'')\Big]\Big\rangle,
\end{align}
where it is left implicit that all the $\hat{T}$ fields inside the integrals are evolved with the free Hamiltonian. We then focus on expanding the nested commutators. Using Wick's theorem, non-vanishing contributions come only from configurations with fully contracted fields, where a contraction between two fields yields:
\beq\label{cont}
    \left\langle\hat{T}_{\vec{q}}(\tau)\hat{T}_{\vec{q}'}(\tau')\right\rangle=\delta(\vec{q}+\vec{q}')\,T_q(\tau)T^*_{q}(\tau'),
\eeq
$T_q$ being the mode function that solves the free equation of motion \eqref{teq}. We note that inverting the order of the two fields in \eqref{cont} gives the complex conjugate of the original contraction. Considering the connected diagrams, where each external leg (labelled by $\vec{k}$ or $\vec{k}'$) should be contracted with a different vertex, one can rewrite the commutators in \eqref{deltaTT} as
\beq \label{comm1}
    \left\langle \left[\left[\hat{T}_{k}\hat{T}_{k'},\hat{T}_{1}\hat{T}_{2}\hat{T}_{3}\right],\hat{T}_{4}\hat{T}_{5}\hat{T}_{6}\right]\right\rangle=\left\langle \left[\left[\hat{T}_{k},\hat{T}_{1}\right]\hat{T}_{k'}\hat{T}_{2}\hat{T}_{3},\hat{T}_{4}\hat{T}_{5}\hat{T}_{6}\right]\right\rangle+\rm{5 \ perms},
\eeq
where the 6 permutations correspond to the contraction of the external $\hat{T}_{k}$ with one of the six internal fields. Note that, in the expression above, we kept the time dependence of the mode functions implicit, as it can be easily recovered from the momentum labels. Before moving on, we find it is useful to count all the possible contractions. The first external leg can be contracted in 6 different ways, while the second external leg has to contract with the opposite vertex, with 3 possible options. Finally, the four internal fields that are left have to be contracted with the two vertices, which can be done in 2 different ways. In total we count $6\times3\times2=36$ different contractions. This allows us to rewrite \eqref{comm1} as
\begin{align}\label{NCs}
   \nn \left\langle \left[\left[\hat{T}_{k}\hat{T}_{k'},\hat{T}_{1}\hat{T}_{2}\hat{T}_{3}\right],\hat{T}_{4}\hat{T}_{5}\hat{T}_{6}\right]\right\rangle &=\delta_{k1}\delta_{k'6}\delta_{24}\delta_{35}\left[T_k(\tau)T_k^*(\tau')-\rm{c.c.}\right]\\ \nn
    &\quad\quad\quad\quad\quad\times\left[T_{k'}(\tau)T_{k'}^*(\tau'')T_{q_2}(\tau')T_{q_2}^*(\tau'')T_{q_3}(\tau')T_{q_3}^*(\tau'')-\rm{c.c.}\right] \\ 
    &\quad +\rm{35 \ perms},
\end{align}
where we reported explicitly the term corresponding to the contractions $\langle k1\rangle$, $\langle k'6\rangle$, $\langle24\rangle$, $\langle35\rangle$ and introduced the notation $\delta_{k1}\equiv\delta(\vec{k}+\vec{q}_1)$. At this point we notice that the integrand in \eqref{deltaTT} is manifestly symmetric under the permutations of possible contractions between the sets $\{\vec{q}_1,\vec{q}_2,\vec{q}_3\}$ and $\{\vec{q}_4,\vec{q}_5,\vec{q}_6\}$, which account for 18 identical contributions. Furthermore, exchanging the two sets $\{\vec{q}_1,\vec{q}_2,\vec{q}_3\}\leftrightarrow\{\vec{q}_4,\vec{q}_5,\vec{q}_6\}$ is equivalent to an exchange of the external momenta $k \leftrightarrow k'$, which gives an extra factor of 2. This leaves us with one permutation with a multiplicity factor of 36. Plugging \eqref{NCs} back in the in-in formula one finds
\begin{align}\label{TTinapp}
    \Big\langle \hat{T}_{\vec{k}}^{(1)}(\tau)\, \hat{T}_{\vec{k}'}^{(1)}(\tau) \Big\rangle' &=-4g^2 \int\,\frac{\d^3q\,\d^3q'}{(2\pi)^{3}}\,\delta(\vec{k}-\vec{q}-\vec{q}') \int^\tau\d\tau'\int^{\tau'}\d\tau'' \\ \nn
    &\quad\quad\times\left[\epsilon_1+\frac{1+m_Q^2}{m_Q\tau'}\epsilon_2+\frac{3m_Q}{\tau'}\epsilon_3\right]_{-\vec{k},\vec{q},\vec{q}\,'}\left[\epsilon_1+\frac{1+m_Q^2}{m_Q\tau''}\epsilon_2+\frac{3m_Q}{\tau''}\epsilon_3\right]_{\vec{k},-\vec{q},-\vec{q}\,'} \\ \nn
    &\quad\quad\times\left[T_k(\tau)T_k^*(\tau')-\rm{c.c.}\right]\left[T_{k}(\tau)T_{k}^*(\tau'')T_{q}(\tau')T_{q}^*(\tau'')T_{q'}(\tau')T_{q'}^*(\tau'')-\rm{c.c.}\right],
\end{align}
where we made the replacements $q_2\rightarrow q$ and $q_3\rightarrow q'$. 
We notice that the last line of \eqref{TTinapp} can be rewritten as
\begin{align}\label{mCs}
    \left[T_k(\tau)T_k^*(\tau')-\rm{c.c.}\right]\left[\dots\right]&= 2i{\rm Im}\left[T_k(\tau)T_k^*(\tau')\right]2i{\rm Im}\left[T_k(\tau)T_k^*(\tau'')\right]{\rm Re}\left[T_{q}(\tau')T_{q}^*(\tau'')T_{q'}(\tau')T_{q'}^*(\tau'')\right] \\ \nn
    &+2i{\rm Im}\left[T_k(\tau)T_k^*(\tau')\right]2i{\rm Re}\left[T_k(\tau)T_k^*(\tau'')\right]{\rm Im}\left[T_{q}(\tau')T_{q}^*(\tau'')T_{q'}(\tau')T_{q'}^*(\tau'')\right].
\end{align}
The \emph{first line} of \eqref{mCs} is symmetric under the exchange $\tau'\leftrightarrow\tau''$, so that the time integration domain can be made rectangular, while accounting for an extra factor $1/2$. This further allows to replace ${\rm Re}[T^4]$ with $T^4$ in the first line of \eqref{mCs}, since its imaginary part ${\rm Im}[T^4]$ is anti-symmetric in $\tau' \leftrightarrow \tau''$  and gives a vanishing contribution to the double integral with (rectangular) symmetric domain in $\tau'$ and $\tau''$. Taking this into account, we write
\begin{align}\label{mC1}
    -4{\rm Im}\left[T_k(\tau)T_k^*(\tau')\right] {\rm Im}[\dots]{\rm Re}[\dots] &= -\mathcal{G}^{(k)}_T(\tau,\tau')\mathcal{G}^{(k)}_T(\tau,\tau'') \left[T_q(\tau')T^*_q(\tau'')T_{q'}(\tau')T^*_{q'}(\tau'')\right],
\end{align}
which has now a separable structure in $\tau'$ and $\tau''$. This leads to the contribution \eqref{P1ttt} in the main text, while the second line of \eqref{mCs} gives \eqref{P2ttt}. \\
Finally, recalling the definitions \eqref{dPS}-\eqref{pertcf}, one can derive the function appearing in Eq.~\eqref{Rs} as the sum of two contributions
\beq\label{f2tttapp}
f_{2,ttt}(x,m_Q) = \sum_{i = 1}^{2} f^{(i)}_{2,ttt}(x,m_Q)\,,
\eeq
where
\begin{align}\label{f2tttapp1}\frac{f^{(1)}_{2,ttt}(x,m_Q)}{{\rm e}^{3\pi (2m_Q + m_Q^{-1})}} &=\frac{1}{2}\int_0^\infty \d \tilde{q} \int_{-1}^{1} \d \eta \,\, \fr{\tilde{q}}{|\hat{k}-\vec{\tilde{q}}|}\Bigg|\int_{x}^{x_{\rm cut}} \d x'\, \fr{{\rm Im}[W_{\alpha,\beta}^{*}(-2ix)W_{\alpha,\beta}(-2ix')]}{W_{\alpha,\beta}(-2ix)} \\ \nn
&\quad\quad \times\left[\tilde{\epsilon}_1 - \frac{1+m_Q^2}{m_Q x'}\epsilon_2-\frac{3m_Q}{x'}\epsilon_3\right] W_{\alpha,\beta}(-2i\tilde{q}x')\,W_{\alpha,\beta}(-2i|\hat{k}-\vec{\tilde{q}}|x')\Bigg|^2,
\end{align}
\begin{align}\label{f2tttapp2}
\frac{f^{(2)}_{2,ttt}(x,m_Q)}{{\rm e}^{3\pi (2m_Q + m_Q^{-1})}} &=\int_0^\infty\d \tilde{q}\int_{-1}^{1} \d \eta\frac{\tilde{q}}{|\hat{k}-\vec{\tilde{q}}|} \int_{x}^{x_{\rm cut}}\d x'\int_{x'}^{x_{\rm cut}} \d x''\,\left[\tilde{\epsilon}_1 - \frac{1+m_Q^2}{m_Q x'}\epsilon_2-\frac{3m_Q}{x'}\epsilon_3\right] \\ \nn
&\quad\quad \times \left[\tilde{\epsilon}_1 - \frac{1+m_Q^2}{m_Q x''}\epsilon_2-\frac{3m_Q}{x''}\epsilon_3\right]^*\fr{{\rm Im}[W_{\alpha,\beta}(-2ix)W_{\alpha,\beta}^*(-2ix')]}{|W_{\alpha,\beta}(-2ix)|^2}\\ \nn
&\quad\quad \times{\rm Re}[W_{\alpha,\beta}(-2ix)W_{\alpha,\beta}^*(-2ix'')] \\ \nn
&\quad\quad \times {\rm Im}[W_{\alpha,\beta}(-2i\tilde{q}x')W_{\alpha,\beta}^*(-2i\tilde{q}x'')\,W_{\alpha,\beta}(-2i|\hat{k}-\vec{\tilde{q}}|x')W_{\alpha,\beta}^*(-2i|\hat{k}-\vec{\tilde{q}}|x'')]\,.
\end{align}
Here we introduced the variable $x=-k\tau$ and the definitions $\tilde{\epsilon}_1\equiv\epsilon_1/k$, $\tilde{q}\equiv q/k$ and $|\hat{k}-\vec{\tilde{q}}| \equiv \sqrt{1-2\tilde{q}\eta+\tilde{q}^2}$. To numerically compute these contributions, for given $m_Q$ values, we use the expressions in Eq.~\eqref{eppapp} for the product of polarization vectors $\tilde{\epsilon}_1,\epsilon_2, \epsilon_3$ provided in terms of the cosine angle $\eta$ and $\tilde{q}$. Moreover, we introduce a cutoff in the upper extremum of integration in $x'$ and $x''$, as defined in \eqref{cutoff}, in order to factor out the sub-horizon oscillatory behaviour of the mode.

\subsection*{Gauge field renormalization by the \emph{tts} interaction}
At cubic order in perturbation theory, tensor fluctuations of the non-Abelian gauge field take up an additional sourced contribution from the tensor-tensor-scalar (\emph{tts}) interaction. We now consider the interaction Hamiltonian \eqref{Htts}, which in Fourier space reads
\beq
\begin{aligned}\label{Httsft}
\hat{H}_{tts}^{(3)}(\tau)
&=-\frac{\lambda}{f}\int\frac{\d^3q_1\d^3q_2\d^3q_3}{(2\pi)^{3/2}}\delta(\vec{q}_1+\vec{q}_2+\vec{q}_3)\left[\vec{\epsilon}(\hat{q}_2)\cdot\vec{\epsilon}(\hat{q}_3)\right]^2 \\
&\times\hat{\delta\chi}(\vec{q}_1)\left\{\frac{g}{2}(aQ\,\hat{T}_+(\vec{q}_2)\hat{T}_+(\vec{q}_3))'+\left[-q_3+(q_3-q_2)\frac{g^2a^2Q^2}{q_1^2+2g^2a^2Q^2}\right]\hat{T}_+'(\vec{q}_2)\hat{T}_+(\vec{q}_3)\right\}\,,
\end{aligned}
\eeq
where the axion perturbation is also promoted to a quantum operator
\beq
\delta \chi(\tau, \vec{x}) = \int \frac{\d^3 q}{(2\pi)^{3/2}}\, {\rm e}^{i \vec{k}.\vec{x}}\, \delta\hat{\chi}(\tau,\vec{q}),\quad\quad \delta\hat{\chi}(\tau,\vec{q})=\delta\chi(\tau, q)\,\hat{a}_\chi(\vec{q})+\delta\chi^*(\tau,q)\,\hat{a}^\dagger_\chi(-\vec{q})\,.
\eeq
As we are going to see, it is convenient to symmetrize over $q_2\leftrightarrow q_3$, so that we get
\begin{align}\label{Httsftsym}
\nn \hat{H}_{tts}^{(3)}(\tau)
& =-\frac{\lambda}{2f}\int\frac{\d^3q_1\d^3q_2\d^3q_3}{(2\pi)^{3/2}}\delta(\vec{q}_1+\vec{q}_2+\vec{q}_3)\left[\vec{\epsilon}(\hat{q}_2)\cdot\vec{\epsilon}(\hat{q}_3)\right]^2\\ \nn 
&\quad\quad\times\Bigg\{\frac{m_Q}{\tau^2}-\left[\frac{m_Q}{\tau}+q_3+\frac{(q_2-q_3)\,m_Q^2}{q_1^2\tau^2+2m_Q^2}\right]\partial_{\tau_1} -\left[\frac{m_Q}{\tau}+q_2+\frac{(q_3-q_2)\,m_Q^2}{q_1^2\tau^2+2m_Q^2}\right]\partial_{\tau_2}\Bigg\} \\
&\quad\quad \times {\delta\hat{\chi}}(\tau,\vec{q}_1)\hat{T}_+ (\tau_1,\vec{q}_2)\hat{T}_+(\tau_2,\vec{q}_3)\Big|_{\tau_1=\tau_2=\tau}. 
\end{align}
Introducing the short-hand notation $\delta(\vec{p}_i + \vec{p}_j + \vec{p}_k) = \delta_{ijk}$, we then plug the resulting interaction Hamiltonian in the in-in formula \eqref{TTin} and obtain
\begin{align}\label{deltaTTtts}
\nn \delta^{(1)}_{(tts)}\Big\langle\hat{T}_{\vec{k}}(\tau)\, \hat{T}_{\vec{k}'}(\tau)\Big\rangle &=-\fr{\lambda^2H^2}{4f^2F_g}\int \prod_{i = 1}^{6} \frac{\d^3q_i}{(2\pi)^{3}}\, \delta_{123} \delta_{345}
\times\left[\vec{\epsilon}(\hat{q}_2)\cdot\vec{\epsilon}(\hat{q}_3)\right]^2\left[\vec{\epsilon}(\hat{q}_5)\cdot\vec{\epsilon}(\hat{q}_6)\right]^2 \int^{\tau}_{\tau_{\rm cut}}\d\tau'\int^{\tau'}_{\tau_{\rm cut}}\d\tau'' \\  \nn 
&\times\Bigg\{\frac{m_Q}{\tau'}-\left[\frac{m_Q}{\tau'}+q_3+\frac{(q_2-q_3)\,m_Q^2}{q_1^2\tau'^2+2m_Q^2}\right]\tau'\partial_{\tau_1'} -\left[\frac{m_Q}{\tau'}+q_2+\frac{(q_3-q_2)\,m_Q^2}{q_1^2\tau'^2+2m_Q^2}\right]\tau'\partial_{\tau_2'}\Bigg\} \\ \nn
&\times\Bigg\{\frac{m_Q}{\tau''}-\left[\frac{m_Q}{\tau''}+q_6+\frac{(q_5-q_6)\,m_Q^2}{q_1^2\tau''^2+2m_Q^2}\right]\tau''\partial_{\tau_1''} -\left[\frac{m_Q}{\tau''}+q_5+\frac{(q_6-q_5)\,m_Q^2}{q_1^2\tau''^2+2m_Q^2}\right]\tau''\partial_{\tau_2''}\Bigg\}\\
&\times\Big\langle \Big[\left[\hat{T}_{\vec{k}}(\tau)\hat{T}_{\vec{k}'}(\tau),\hat{X}_{\vec{q}_1}(\tau')\hat{T}_{\vec{q}_2}(\tau_1')\hat{T}_{\vec{q}_3}(\tau_2')\right],\hat{X}_{\vec{q}_4}(\tau'')\hat{T}_{\vec{q}_5}(\tau_1'')\hat{T}_{\vec{q}_6}(\tau_2'')\Big]\Big\rangle\Big|_{\tau_1'=\tau_2'=\tau',\tau_1''=\tau_2''=\tau''},
\end{align}
where we introduced the canonically-normalized axion perturbation $\hat{X}=a(\tau)\sqrt{F_g}{\delta\hat{\chi}}\simeq \sqrt{F_g}\delta\hat{\chi}/(-H \tau)$, with $F_g$ treated as a constant as detailed in Appendix \ref{AppB}. \\
The nested commutators are expanded as follows
\beq
\begin{aligned}
\left\langle \left[\left[\hat{T}_{k}\hat{T}_{k'},\hat{X}_{1}\hat{T}_{2}\hat{T}_{3}\right],\hat{X}_{4}\hat{T}_{5}\hat{T}_{6}\right]\right\rangle &=\delta_{k2}\delta_{k'5}\delta_{36}\delta_{14}\left[T_k(\tau)T_k^*(\tau_1')-\rm{c.c.}\right] \\
&\times\left[T_{k'}(\tau)T_{k'}^*(\tau_1'')X_{q_1}(\tau')X_{q_1}^*(\tau'')T_{q_3}(\tau_2')T_{q_3}^*(\tau_2'')-\rm{c.c.}\right]+\rm{7 \ perms},
\end{aligned}
\eeq
where the different permutations come from the possible contractions of the $\hat{T}_+$ fields, accounting only for connected configurations. By symmetry, the different permutations of the nested commutators amount to the same contribution, giving an overall factor of 8. After relabeling the internal momenta, $q_3=q_6\equiv q$ and $q_1=q_4\equiv q'$,  one arrives  at Eq.~\eqref{P1tts} reported in the main text, where we used the definition \eqref{dPS}. \\
Recalling the definition \eqref{pertcf}, and using the variable $x\equiv-k\tau$, one obtains the function $f_{2,tts}$ in Eq.~\eqref{Rs} as
\begin{align}\label{f2ttsapp}
\frac{f_{2,tts}(x,m_Q)}{{\rm e}^{2\pi (2m_Q + m_Q^{-1})}} &=\frac{\lambda^2H^2}{64\pi^2f^2F_g}\int_0^\infty\d \tilde{q}\int_{-1}^{1} \d \eta\frac{\tilde{q}(1+\eta)^4}{|\hat{k}-\vec{\tilde{q}}|}\int_{x}^{x_{\rm cut}}\d x'\int_{x'}^{x_{\rm cut}} \d x''  \\ \nn  
& \quad\quad\quad\quad\quad\times\mathcal{\tilde{D}}_{x_1'x_2'}\,\mathcal{\tilde{D}}_{x_1''x_2''}\,\frac{{\rm Im}\left[W_{\alpha,\beta}(-2ix)W_{\alpha,\beta}^*(-2ix_1')\right]}{\left|W_{\alpha,\beta}(-2ix)\right|^2} \\ \nn 
&\quad\quad\quad\quad\quad\times{\rm Im}\Big[W_{\alpha,\beta}(-2ix)W_{\alpha,\beta}^*(-2ix_1'')W_{\alpha,\beta}(-2i\tilde{q}x_2')W_{\alpha,\beta}^*(-2i\tilde{q}x_2'') \\ \nn &\quad\quad\quad\quad\quad\times X_c(|\hat{k}-\vec{\tilde{q}}|x')X_c^*(|\hat{k}-\vec{\tilde{q}}|x'')\Big]\Big|_{x_1'=x_2'=x', x_1''=x_2''=x''},
\end{align}
with $\tilde{q}\equiv q/k$, $|\hat{k}-\vec{\tilde{q}}| \equiv \sqrt{1-2\tilde{q}\eta+\tilde{q}^2}$, $X_c\equiv\sqrt{2k}X$. The operator $\mathcal{\tilde{D}}_{x_1x_2} \equiv -\mathcal{D}_{\tau_1\tau_2}/k$ acting on the gauge tensors can be rewritten as
\beq
\mathcal{\tilde{D}}_{x_1x_2}\equiv \frac{m_Q}{x} -\left[\frac{m_Q}{x}-\tilde{q}+\frac{(\tilde{q}-1)\,m_Q^2}{|\hat{k}-\vec{\tilde{q}}|^2x^2+2m_Q^2}\right]x\partial_{x_1}-\left[\frac{m_Q}{x}-1+\frac{(1-\tilde{q})\,m_Q^2}{|\hat{k}-\vec{\tilde{q}}|^2x^2+2m_Q^2}\right]x\partial_{x_2}.
\eeq
The explicit expression for $x_{\rm cut}$ is given in Eq.~\eqref{cutoff}. 
\medskip

\noindent{\bf Cut-off procedure for the computation of loop integrals.} The integrals \eqref{f2tttapp} (see Eqs.~\eqref{f2tttapp1} and \eqref{f2tttapp2}) and \eqref{f2ttsapp} appearing in the loop corrections are divergent as they take up contributions from the non-physical vacuum modes of the tensor and axion fluctuations. One would isolate these contributions by restricting the upper limit of the time integration over the internal $x'$ to the time $x_{\rm th}$ when the tachyonic instability for the tensor gauge mode begins (Eq.~\eqref{xinsta}). However this choice is not sufficient for eliminating the unphysical contributions as the arguments of the field fluctuations are re-scaled by factors of $\tilde{q}$ and $|\hat{k}-\vec{\tilde{q}}| = \sqrt{1-2\eta \tilde{q} + \tilde{q}^2}$, implying that the integration over a given internal loop momentum configuration $\{\eta,\tilde{q}$\} would still receive contributions from the vacuum modes. A more proper way to isolate the vacuum fluctuations is to use an adjusted cut-off that is a function of $\{\eta,\tilde{q}$\} 
\beq\label{cutoff}
x_{\rm cut} =  x^{(+)}_{\rm th}\times\min\left[1,\frac{1}{\tilde{q}},\frac{1}{|\hat{k}-\vec{\tilde{q}}|}\right],
\eeq
where $x^{+}_{\rm th}$ (Eq.~\eqref{xinsta}) denotes the onset of instability for the gauge tensor modes. The cut-off \eqref{cutoff} ensures that the integration starts only after when all the Whittaker's functions that appear in the integrands of \eqref{f2tttapp1},\eqref{f2tttapp2} and \eqref{f2ttsapp} are out of their vacuum configuration.
\medskip

\noindent{\bf Polarization vectors and their products.}
Let us provide some definitions and details regarding the product of polarization vectors that appear in the one-loop correlators we studied above. We begin by noting the general expression of the polarization vectors in spherical coordinates:
\begin{equation}\label{eg}
    \vec{\epsilon}_{\pm}(\vec{q})=\frac{1}{\sqrt{2}}\left(\cos\theta\cos\phi\mp i\sin\phi,\cos\theta\sin\phi\pm i\cos\phi,-\sin\theta\right),
\end{equation}
for an arbitrary vector argument $\vec{q}$ with the following components
\begin{equation}\label{vg}
\vec{q}=q\left(\sin\theta\cos\phi,\sin\theta\sin\phi,\cos\theta\right).
\end{equation}
The two-point correlators we work with involve products of two or more polarization vectors that includes the external momenta $\vec{k}$ in their arguments. Since we are interested in one-loop computations including one internal momenta, we have the freedom to align the external $\vec{k}$ along a preferred direction, which we choose to be the $z$ axis, corresponding to the choices $\theta=0$ and $\phi=0$ in \eqref{vg} and \eqref{eg}:
\begin{equation}
    \vec{\epsilon}_{\pm}(\vec{k})
    =\frac{1}{\sqrt{2}}\left(1,\pm i,0\right), \quad\quad
    \vec{k}
    =k\left(0,0,1\right).
\end{equation}
Due to the azimuthal symmetry, we can then choose the internal momentum $\vec{q}$ to lie in the $x-z$ plane, corresponding to $\phi=0$:
\begin{equation}
    \vec{\epsilon}_{\pm}(\vec{q})
    =\frac{1}{\sqrt{2}}\left(\eta,\pm i,-\sqrt{1-\eta^2}\right), \quad\quad
    \vec{q}
    =q\left(\sqrt{1-\eta^2},0,\eta\right),
\end{equation}
where the cosine angle between $\vec{k}$ and $\vec{q}$ is given by $\hat{k}\cdot\hat{q}=\cos\theta\equiv\eta$. From \eqref{vg} and \eqref{eg}, one can also infer a general expression for the polarization vectors in terms of the components of its argument vector as 
\beq
\epsilon_{\pm}(\vec{q}) = \frac{1}{\sqrt{2}}\left(\frac{q_x q_z \pm i q_y|\vec{q}|}{|\vec{q}| \sqrt{q_y^2+q_z^2}},\frac{q_x q_y\mp i q_z|\vec{q}|}{|\vec{q}| \sqrt{q_y^2+q_z^2}}, -\frac{\sqrt{q_y^2+q_z^2}}{|\vec{q}|}\right) .
\eeq
Following these relations, the products of polarization vectors \eqref{ept} can be obtained as 
\begin{align}\label{eppapp}
\nn \tilde{\epsilon}_1(-\vec{k},\vec{q},\vec{k}-\vec{q}) &= -\frac{\left(1-\eta^2\right)}{8} \left(1+\frac{1+\tilde{q}}{\left(1-2 \tilde{q} \eta+\tilde{q}^2\right)^{1 / 2}}\right)\\ \nn
& \quad\quad\quad\quad\quad\quad \times \left[\tilde{q} \left(1+\frac{\tilde{q}-\eta}{\left(1-2 \tilde{q} \eta+\tilde{q}^2\right)^{1 / 2}}\right)+\left(1+\frac{1+\tilde{q}}{(1+2\tilde{q}\eta+\tilde{q}^2)^{1/2}}\right)\right], \\ \nn
\epsilon_2(-\vec{k},\vec{q},\vec{k}-\vec{q}) &= -\frac{\left(1-\eta^2\right)}{8}\left(1+\frac{1+\tilde{q}}{\left(1-2 \tilde{q} \eta+\tilde{q}^2\right)^{1 / 2}}\right)^2, \\
\epsilon_3(-\vec{k},\vec{q},\vec{k}-\vec{q}) &= -\frac{(1+\eta)}{8}\left(1+\frac{1-\tilde{q} \eta}{\left(1-2 \tilde{q} \eta+\tilde{q}^2\right)^{1 / 2}}\right)\left(1+\frac{\tilde{q}-\eta}{\left(1-2 \tilde{q} \eta+\tilde{q}^2\right)^{1 / 2}}\right).
\end{align}

\section{Dynamics of axion perturbations for \emph{tts} contribution}\label{AppB}

In this Appendix, we study the dynamics of the axion perturbations, necessary to obtain leading-order \emph{tts} loop contributions to the gauge field wave-function (see \eg \eqref{f2ttsapp}). \\
\indent The evolution of the perturbation $\delta\chi$ in the axion sector is highly sensitive to the main source of friction that slows down the axion vev at the background level, leading to a rich set of possibilities for the dynamics of $\delta \chi$. In this work, we focused on the recent model studied in \cite{Dimastrogiovanni:2023oid} where the axion-inflaton has an additional (w.r.t. the friction provided by the gauge sector) source of friction through a non-minimal coupling with gravity:
\beq\label{Lnm}
\frac{\mathcal{L}_{\rm NM}}{\sqrt{-g}} = -\frac{1}{2}\left(g^{\mu\nu} -  \frac{G^{\mu\nu}}{M^2}\right)  \partial_\mu \chi \partial_\nu \chi - V(\chi)\,.
\eeq
Here $V(\chi)$ is the (shift-symmetric) axion potential and $G_{\mu\nu}$ the Einstein tensor. For a sufficiently large non-minimal coupling, at the initial stages of inflation the background settles into a slow-roll trajectory in a gravitationally-enhanced friction regime, analogous to the one of UV-protected inflation \cite{Germani:2010hd}. With the addition of this non-minimal coupling, the model has been shown to produce viable predictions for CMB observables \cite{Dimastrogiovanni:2023oid}.\\
\indent In the following, we study the dynamics of the scalar perturbations in the so called gravitationally-enhanced friction (GEF) regime, which can be parametrized in terms of the background quantities locked in the slow-roll trajectory as 
\beq\label{gefc}
  \frac{m_Q^2 \Lambda^2}{3 F_g}\ll 1, \quad\quad \fr{\Lambda^2}{2 F_g}\ll 1,
\eeq
having defined the dimensionless parameters $m_Q \equiv g Q /H$, $\Lambda \equiv \lambda Q /f$. Notice that the ratios \eqref{gefc} are suppressed by the factor $F_g$ that parametrizes the strength of the non-minimal coupling between the axion-inflaton and gravity: 
\beq
F_g = 1 + \frac{3H^2}{M^2} \gg 1 \quad\quad \longleftrightarrow \quad\quad \frac{H}{M} \gg 1,
\eeq
where in the slow-roll trajectory $
{\dot{F_g}}/{(H\,F_g)}\ll 1$ is satisfied, so that  $ F_g \simeq \textrm{constant}$.
The dynamics of scalar perturbations in the GEF regime is studied in details in \cite{Dimastrogiovanni:2023oid}. In the following discussion, we consider the same coupled scalar system adopting the conventions from \cite{Adshead:2013nka,Papageorgiou:2018rfx}. 

In particular, we work in the decoupling limit, ignoring the non-dynamical scalar modes of the metric in the spatially flat gauge $\delta g_{ij,{\, \rm scalar}} = 0$. The full scalar system of perturbations contains the inflaton fluctuation $\delta \chi$ together with the scalar fluctuations $\delta\phi, Z, \chi_3, \delta A^{3}_0$ of the ${\rm SU}(2)$ sector. Without loss of generality, we align the Fourier momenta of the perturbations along the z-axis to describe the scalar components of the ${\rm SU}(2)$ multiplet as
\begin{align}\label{GPert}
\nn \delta A_\mu^1 & =\left(0,\, \delta \phi(\tau, z)-Z(\tau, z),\, \chi_3(\tau, z),\,0\right)\,, \\
\nn \delta A_\mu^2 & =\left(0,\,-\chi_3(\tau, z),\, \delta \phi(\tau, z)-Z(\tau,z),\, 0\right)\,, \\
\delta A_\mu^3 & =\left(\delta A_0^3(\tau, z),\, 0,\, 0,\, \delta \phi(\tau,z)+2 Z(\tau, z)\right),
\end{align}
where $\chi_3$ can be eliminated in favor of  $\delta \phi$ and $Z$ by fixing the gauge in the ${\rm SU}(2)$ sector:
\beq\label{chi3}
\chi_3=-i k \frac{2 Z+\delta \phi}{2 g a Q}.
\eeq
Using \eqref{GPert} and \eqref{chi3} in \eqref{LAGF}, one can notice that $\delta A_0^{3}$ is also a non-dynamical mode and it can be expressed in terms of $Z$ and $\delta \phi$ as
\beq\label{dA03}
\delta A_{0}^3 = -i k\, \frac{g a^2 Q^2 \lambda\, \delta \chi / f - 2\, (2Z + \delta \phi) (a Q)^{\prime}/(aQ)}{k^2+2 g^2 a^2 Q^2}.
\eeq
Eliminating $\delta A_{0}^3$ with \eqref{dA03}, we are then left with a scalar system of perturbations described by $\delta \chi, \delta \phi$ and $Z$. Notice that, since the non-minimal coupling does not influence the gauge sector, the expressions we found in \eqref{chi3} and \eqref{dA03} are identical to the ones presented in \cite{Papageorgiou:2018rfx}. On the other hand, the non-minimal coupling \eqref{Lnm} determines the canonical normalization of the axion perturbation. Inspecting the second order action that arises from the gauge sector \eqref{LAGF}, one can infer that $\delta \phi$ and $Z$ are not canonically normalized. To study the system, we diagonalize the kinetic terms using the following canonical variables: 
\beq\label{cnsp}
\hat{X} = a(\tau)\sqrt{F_g}\,\delta\chi, \quad \hat{Z}=\sqrt{2}(Z-\delta \phi), \quad \hat{\varphi} \equiv \sqrt{2+\frac{x^2}{m_Q^2}}\left(\frac{\delta \phi}{\sqrt{2}}+\sqrt{2} Z\right)\,.
\eeq
In terms of the canonical variables, the equations of motion of the physical scalar perturbations read  
\begin{align}\label{Xeq}
    &\nn \hat{X}'' + \left[1-\frac{2}{x^2}+\fr{1}{F_g}\left(\frac{\Lambda^2m_Q^2 }{2m_Q^2+x^2}+\frac{V''}{H^2x^2}\right)\right]\hat{X} + \left(\frac{\sqrt{2}\Lambda(4m_Q^4+3m_Q^2x^2+x^4)}{x^2(2m_Q^2+x^2)^{3/2}}\right)\fr{\hat{\varphi}}{\sqrt{F_g}} \\ 
    &\quad\,\, - \frac{2\sqrt{2}\Lambda m_Q}{x^2}\fr{\hat{Z}}{\sqrt{F_g}} - \frac{\sqrt{2}\Lambda m_Q^2}{x \sqrt{2m_Q^2+x^2}}\fr{\hat{\varphi}'}{\sqrt{F_g}} + \frac{\sqrt{2}\Lambda m_Q}{x}\fr{\hat{Z}'}{\sqrt{F_g}} = 0 \,,
\end{align}
\begin{align}\label{vphieq}
    &\nn \hat{\varphi}'' + \left(1-\frac{2}{2m_Q^2+x^2}+\frac{2 m_Q^2}{x^2}+\frac{
   6m_Q^2}{(2m_Q^2+x^2)^2}\right)\hat{\varphi} + \left(\frac{2\sqrt{2m_Q^2+x^2}}{m_Qx^2}\right)\hat{Z} + \frac{\sqrt{2}\Lambda m_Q^2}{x \sqrt{2m_Q^2+x^2}}\fr{\hat{X}'}{\sqrt{F_g}}\\
   &\quad\,\,  + \left(\frac{\sqrt{2}\Lambda(2m_Q^2+ m_Q^2 x^2 + x^4)}{x^2(2m_Q^2+x^2)^{3/2}}\right)\fr{\hat{X}}{\sqrt{F_g}} = 0 \,.
\end{align}
\begin{align}\label{Zeq}
    & \hat{Z}'' + \left(1-\frac{2-2m_Q^2}{x^2}\right)\hat{Z} - \frac{\sqrt{2}\Lambda m_Q}{x^2}\fr{\left(\hat{X}+x\hat{X}'\right)}{\sqrt{F_g}} + \left(\frac{2\sqrt{2m_Q^2+x^2}}{m_Qx^2}\right)\hat{\varphi} = 0,
\end{align}
\begin{figure}[t!]
\begin{center}
\includegraphics[scale=0.55]{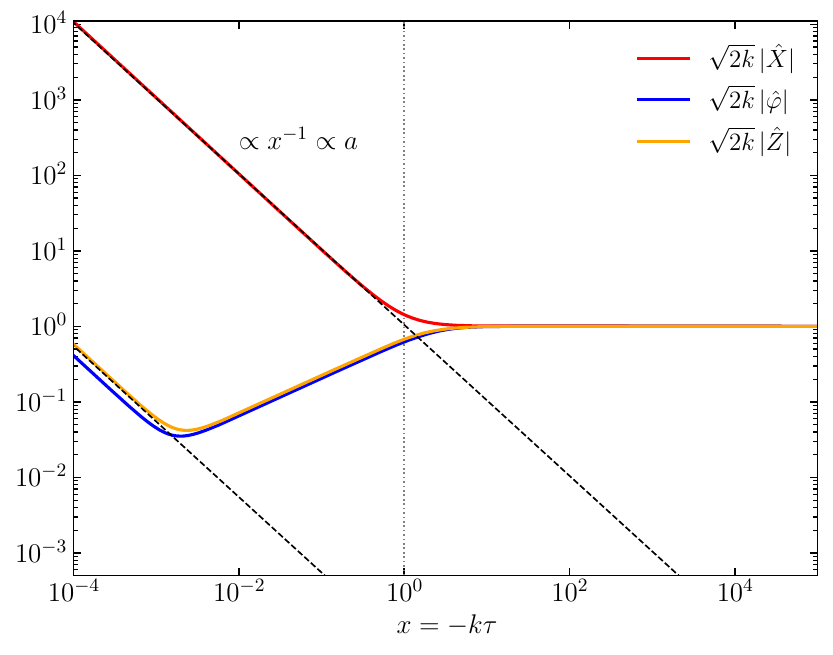}\includegraphics[scale=0.55]{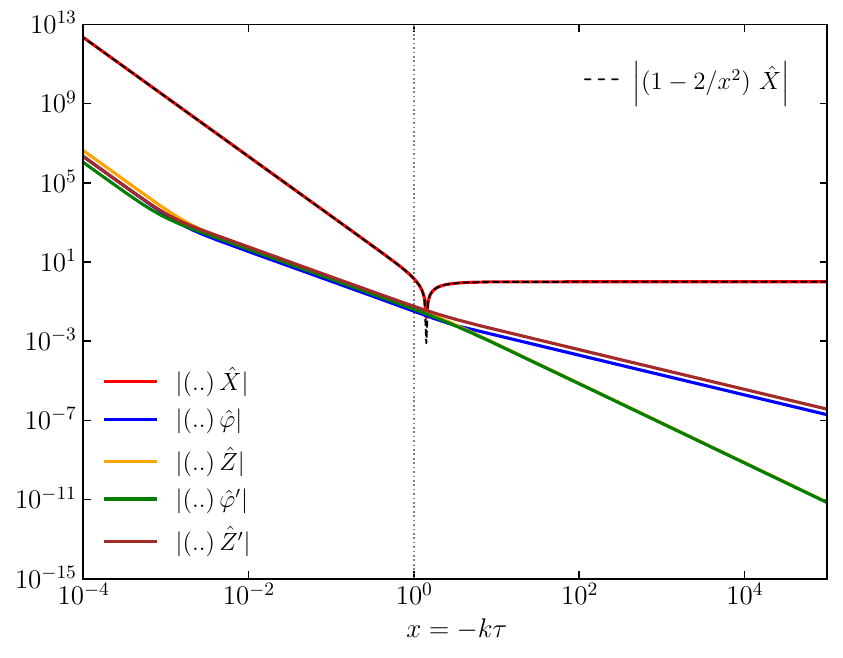}
\end{center}
\vspace*{-5mm}\caption{Left: Evolution of the scalar perturbations \eqref{cnsp}, derived by solving  \eqref{Xeq}-\eqref{Zeq} with BD initial conditions. Diagonal dotted lines $\propto x^{-1} \propto a$ represent the late time behavior of the fluctuations. Right: Contributions of various terms in the axion equation \eqref{Xeq}. In the numerical evaluations, the background quantities are assumed to be frozen at their $N = 60$ values, with $m_Q \simeq 1.9$ and $\Lambda \simeq 0.99$, using the parameters provided in the text. 
\label{fig:scapert}}
\end{figure}

\noindent Focusing on a parameter set $\{g = 0.1, \mu = 0.005, \lambda = 550, f = 0.2, M = 9 \times 10^{-7}\}$ 
that provides a sufficiently long slow-roll evolution in the GEF regime \cite{Dimastrogiovanni:2023oid}, we initialize the multi-field system of scalar fluctuations $\hat{X},\hat{\varphi}$ and $\hat{Z}$ in the Bunch Davies (BD) vacuum to numerically solve the coupled dynamics. The solutions, as the modes evolve from inside to outside the horizon (from right to left), are shown in Fig. \ref{fig:scapert}. We observe from the left panel that, in the GEF regime, the canonical axion perturbation $\hat{X}$ remains decoupled from the rest of the dynamics, while the solutions for $\hat{\varphi}$ and $\hat{Z}$ trace each other's behaviour.
On super-horizon scales, all modes behave as $\propto x^{-1} \propto a$. The decoupling of the canonical axion perturbation is more clearly shown in the right panel, where we plot various source terms in \eqref{Xeq}, obtained through numerical evaluation. In particular, we notice that the diagonal mass term is the dominant contribution to the dynamics of $\hat{X}$. More importantly, we observe that this term can be approximated by the standard expression $\propto (1-2/x^2)$, which prompts us to conclude that in the GEF regime, the axion perturbation evolves as a standard scalar perturbation in a (quasi) de-Sitter background (up to slow-roll suppressed terms):
\beq \label{Xeqapprox}
\hat{X}'' + \left[1-\fr{2}{x^2}\right]\hat{X} \simeq 0.
\eeq
A close inspection of the system \eqref{Xeq}-\eqref{Zeq} confirms these findings, as in the gravitationally enhanced friction regime all the non-diagonal source terms in the equation of motion for the canonical axion perturbation are suppressed by large factors of $F_g \gg 1$. Notice that the same is true for the last contribution to the mass of $\hat{X}$ inside the square brackets. Similarly, mixing terms $\propto \hat{X}, \hat{X}'$ in the equations of motion of $\hat{\varphi}$ and $\hat{Z}$ can be neglected in this regime, leaving out only mixing terms $\propto \hat{Z}$ and $\propto \hat{\varphi}$ respectively. These terms in turn dictate the interdependence of the $\hat{\varphi} - \hat{Z}$ dynamics for $x \lesssim 1 $, as shown in the left panel of Figure \ref{fig:scapert}. In comparison, in the standard CNI setup \cite{Adshead:2013nka}, the axion sector has minimal coupling with gravity and the behavior of axion perturbations around horizon crossing differs with respect to the case we study here (see e.g. Fig.~\ref{fig:scapert} versus Fig.~1 of \cite{Adshead:2013nka})\\
\indent Eq. \eqref{Xeqapprox} implies that in the GEF regime of \cite{Dimastrogiovanni:2023oid}, the solution for the canonical axion perturbations can be accurately described by the well known solution 
\beq \label{Xc}
X_c(x,k) = \sqrt{2k}\, \hat{X}(x) = \left(1+\frac{i}{x}\right){\rm e}^{ix}.
\eeq
In this work, we utilize this expression to compute the resulting $tts$ one-loop diagram, relevant for the perturbativity criterion (see \eg Eq. \eqref{f2ttsapp}). 
\end{appendix}

\addcontentsline{toc}{section}{References}
\bibliographystyle{utphys}
\bibliography{paper2}
\end{document}